%% file: mnras_template.tex
\documentclass[fleqn,usenatbib]{mnras}

\usepackage{newtxtext,newtxmath}

\usepackage[T1]{fontenc}

\DeclareRobustCommand{\VAN}[3]{#2}
\let\VANthebibliography\thebibliography
\def\thebibliography{\DeclareRobustCommand{\VAN}[3]{##3}\VANthebibliography}

\usepackage{graphicx}	
\usepackage{amsmath}	
\usepackage{xspace}
\usepackage[noabbrev]{cleveref}
\usepackage{soul}
\usepackage{subcaption}
\usepackage{widetext}
\usepackage{subcaption}

\newcommand{\dz}{{\rm d}z}

\newcommand{\diff}{{\rm d}}

\newcommand{\lya}{Ly$\alpha$\xspace}
\newcommand{\lyaf}{Ly$\alpha$ forest\xspace}
\newcommand{\nvec}{\ensuremath{\hat{\mathbf{n}}}\xspace}

\definecolor{mypink3}{cmyk}{0, 0.7808, 0.4429, 0.1412}

\title[Clustering redshifts with Lyman-$\alpha$ forest]{Calibrating redshift distributions at $z>2$ with Lyman-$\alpha$ forest cross-correlations}

\input{authors}

\date{Accepted XXX. Received YYY; in original form ZZZ}

\pubyear{\the\year{}}

\begin{document}
\label{firstpage}
\pagerange{\pageref{firstpage}--\pageref{lastpage}}
\maketitle

\begin{abstract}
We explore the feasibility of using Lyman-$\alpha$ (Ly$\alpha$) forests to calibrate the ensemble redshift distribution of the high-redshift tail ($2<z<3$) of photometric galaxies. We use \texttt{CoLoRe} simulations to create mock DESI 5-year Ly$\alpha$ forests and Rubin Observatory LSST 10-year photometric galaxies up to $z=3$, and measure the galaxy redshift distribution via their angular cross-correlations. Due to large redshift-space distortions in the Ly$\alpha$ forest, the conventional $n(z)$ estimator for clustering redshifts does not apply, and we develope a theoretical framework to model the angular cross-correlation directly. {Using the simulations, we explore effects of instrumental noise, continuum fitting, contamination in the Ly$\alpha$ forest, cross-correlation angular scales ($\theta$), and redshift bin size ($\Delta z$) on the signal-to-noise (SNR) of the measurements.} We find that continuum fitting methods strongly impact the SNR of the measurements. With our baseline continuum fitting method, \texttt{LyCAN}, at angular scales $\theta\sim10$ arcmin and $\Delta z=0.1$, we measure the cross-correlation signal at $24\sigma$. If the shape of the redshift distribution and galaxy bias evolution are known well for $z<2$, the cross-correlation can constrain the mean redshift of the galaxy sample to $\sigma_z/(1+\bar{z}) = 0.006$ at a mean redshift of $\bar{z}=2$. This demonstrates that Ly$\alpha$ cross-correlation is a reliable and promising method to calibrate the high-redshift tails of photometric Stage IV galaxy surveys.
\end{abstract}




\section{Introduction} \label{sec:intro}

Cosmological probes such as weak gravitational lensing rely on photometric galaxy surveys. A main source of uncertainty in the analysis is the determination of redshifts of the photometric galaxies. The common practice is to use broadband photometry to provide an estimate of the redshift (or redshift probability density distribution) for each galaxy, called the `photo-$z$', via template fitting 
or machine learning 
\citep[see][and references therein]{Newman_2022}. Ensemble redshift distributions can also be estimated directly, via colour-redshift space clustering methods such as Self-Organizing Maps \citep[][]{2015ApJ...813...53M}. 
However, these methods critically rely on good knowledge of the colour-redshift relation, i.e. either a complete set of SED templates, or a complete spectroscopic training sample. This is increasingly difficult to achieve for faint and distant sources, increasing potential uncertainties and biases in redshift calibration.

An alternative route is called cross-correlation redshifts \citep[or clustering redshifts;][]{Newman_2008, 2013arXiv1303.4722M, Schmidt_2013}, which uses information from the clustering property of galaxies as tracers of the large-scale structure.
By measuring the angular cross-correlation between the photometric \textit{unknown} sample and a spectroscopic \textit{reference} sample split in narrow redshift bins, one can recover the redshift distribution of the unknown sample.
With no dependence on the galaxy photometry for either sample, this method provides orthogonal constraints for the redshift distribution obtained from the photo-$z$, in some cases decreasing the mean uncertainties by a factor of two \citep{2025arXiv251023565D}.
Some recent applications include the Kilo-Degree Survey Legacy data \citep[KiDS-Legacy;][]{2025arXiv250319440W}, the Dark Energy Survey Year-6 data\citep[DES Y6;][]{2025arXiv251023565D}, and Hyper Suprime-Cam Year-3 data \citep[HSC Y3;][]{2023MNRAS.524.5109R, 2025arXiv251118133C, 2025arXiv251215963R}. 
In both KiDS-Legacy and HSC Y3 analyses, the reference samples run out at $z\sim1.2$, leaving their higher redshift tails uncalibrated by the clustering method. DES Y6 included BOSS/eBOSS emission line galaxy (ELG) and quasar (QSO) samples to push the calibration to $z\sim2$. 
Forthcoming stage IV surveys such as Vera C. Rubin Observatory's Legacy Survey of Space and Time \citep[LSST;][]{2019ApJ...873..111I} expect to have a significant fraction of galaxies at $z>2$ in the highest source tomographic bin. Although this redshift range is covered by DESI QSOs up to $z\sim3$, the clustering method can become challenging as the QSO number density drops sharply beyond $z>2$ and effects such as magnification bias complicate the clustering signal.
On the other hand, the unconstrained high-redshift tail introduces uncertainties on the mean redshift of the source galaxy sample, degrading the cosmological parameter constraints \citep[e.g.][]{2023PhRvD.108l3519D}.

A promising $z>2$ reference sample is the Lyman-$\alpha$ (\lya) forest, which refers to a series of absorption features in the continuum of high-redshift quasar spectra caused by neutral hydrogen (HI) in the inter-galactic medium (IGM) at $\lambda_0=1215.67${\AA} in the rest frame. The \lya optical depth is proportional to the baryon density and the HI fraction in the IGM, and the redshift of the gas responsible for the absorption is given by $z = (\lambda - \lambda_0)/\lambda_0$, where $\lambda$ is the observed wavelength. Hence, the \lyaf can be regarded as a tracer of the underlying density field. 
Indeed, the 3D correlation function of the \lyaf was first measured in SDSS \citep{2011JCAP...09..001S}, and subsequently used to measure Baryon Acoustic Oscillations (BAO) by e.g. \cite{2013JCAP...04..026S, 
2013A&A...552A..96B, 
dMdB2020, DESI_DR2_Lya_BAO}. This was also measured using \lya $\times$ QSO \cite[e.g][]{2014JCAP...05..027F}.
Full shape analysis has also been applied \citep[e.g.][]{2023PhRvL.130s1003C} for precision cosmology. These studies show that linear theory well describes the \lyaf correlation function, and provide good measurements of the \lya bias and the RSD parameter on scales larger than $30 h^{-1}$Mpc.

The main challenge of using \lyaf for clustering redshifts is the impact of continuum fitting.
The large-scale structure tracer of interest here is the fluctuation in the transmitted flux fraction, $\delta_F^q (\lambda)$, given a quasar line-of-sight, $q$, as a function of the observed wavelength, $\lambda$. This quantity can be computed using the observed flux, $f^q(\lambda)$, the quasar continuum, $C^q (\lambda)$, and the mean transmitted flux fraction, $\bar{F}(\lambda)$:
\begin{equation}
    \delta_F^q(\lambda) = \frac{f^q(\lambda)}{\bar{F}(\lambda)C^q(\lambda)}-1.
    \label{eq: lya deltaf}
\end{equation}
The function $\bar{F}(\lambda)C^q(\lambda)$ is typically unknown and needs to be estimated from the data.
This is typically done by \texttt{Picca}, the continuum fitting code used in \lya BAO analysis from the eBOSS and DESI collaborations \citep{dMdB2020, DESI_DR1_Lya_BAO, DESI_DR2_Lya_BAO}. However, this procedure suppresses radial modes on large scales, introducing distortions in the $\delta^q_F(\lambda)$ two-point correlation and reducing the signal-to-noise significantly.
This large-scale normalization problem does not exist in \lyaf alone.
A similar study in spirit is \cite{2019MNRAS.482.3341C}, who proposed to use HI intensity mapping as the reference sample for cross-correlation redshift calibration. In this case, the foreground cleaning process of HI intensity maps renders the large-scale radial modes useless for cross-correlation {(although recent studies suggest that those modes can be recovered, see e.g. \citealp{2026PhRvD.113b3521S})}. They tackled the problem by removing these modes in Fourier space and only measuring the correlation on small scales. However, as the foreground removal becomes more aggressive, the signal-to-noise decreases and the bias in the estimated unknown galaxy redshift distribution $n(z)$ increases.

Recently, \cite{2024ApJ...976..143T} introduced a machine learning-based continuum fitting method, \texttt{LyCAN}, that predicts the quasar continuum based on the observed fluxes red-ward of the \lya emission line in the quasar spectrum.
Using only pixels free of \lya absorption, \texttt{LyCAN} does not introduce correlations in the large radial modes of $\delta_F$, making it usable for clustering redshift studies.
In this paper, we investigate, in the context of DESI 5-year and Rubin LSST 10-year data, how well \lyaf can be used in clustering redshift with this new continuum fitting method. We set up the modeling framework for clustering redshift at high-redshift, and investigate the performance varying the level of realism of $\delta_F$,  angular scales, redshift bin size, survey area, and the photometric galaxy number density. 

This paper is organized as follows. We introduce the simulations used in this paper in Section~\ref{sec:simulations}, and we describe the two continuum fitting methods in Section~\ref{sec: Estimating the fluctuations}. Section~\ref{sec: theory} presents the theoretical framework for \lyaf clustering redshifts. Section~\ref{sec: estimators} presents how the cross-correlation is measured, along with models for the bias evolution and redshift distribution, covariance matrix, and likelihoods. Section~\ref{sec: results} shows our baseline results and how the results vary with \lyaf types, the angular scales, and the redshift binning of the reference sample. Finally, we summarize in Section~\ref{sec: conclusions}.



\input{simulations_new}

\input{continuum_fitting_new}

\section{Clustering redshift theory}
\label{sec: theory}

In this section, we derive an expression for the two-point angular cross-correlation function of the binned flux fluctuation, $\Delta_F(\hat{\mathbf{n}}, z_I)$ and the unknown galaxy number density $\Delta_{g}(\hat{\mathbf{n}})$:
\begin{equation}
    w_{gF}(\theta,z_I) = \langle \Delta_F(\hat{\mathbf{n}},z_I)\Delta_g(\hat{\mathbf{n}}') \rangle,
\end{equation}
where $\theta = |\hat{\mathbf{n}} - \hat{\mathbf{n}}'|$ and $\langle...\rangle$ represents spatial average over all $\hat{\mathbf{n}}, \hat{\mathbf{n}}'$ over the survey footprint.

We give an overview of the conventional clustering redshift estimator and its main assumptions in Appendix~\ref{sec: old cc}. Briefly, the conventional estimator for the unknown redshift distribution is constructed by $n_{\rm u}(z)\propto w_{\rm ru}/\sqrt{w_{\rm rr}}$, where $w_{\rm ru}$ is the angular cross-correlation between the reference and unknown sample, and $w_{\rm rr}$ is the auto-correlation of the reference sample.
The expression is valid based on the following assumptions: 1) redshift-space distortion and magnification for the auto- and cross-correlations can be ignored;
2) contributions to the correlation function only come from pairs at the same redshift (Limber approximation). Both of these are essentially `edge effects', which are negligible when the correlation is measured on a scale much smaller than the redshift bin width. The typical low-redshift clustering method takes a bin width of order $\Delta z\sim 0.05$, and the correlation is measured on a scale corresponding to $\mathcal{O}(1)h^{-1}$Mpc.
However, at a fixed redshift bin width $\Delta z$, the corresponding radial comoving width at $z>2$ is smaller compared to lower redshifts, while the transverse comoving scale at fixed angle becomes larger. 
This makes the edge effects important and the redshift distribution estimator becomes biased (see Appendix~\ref{sec: Modeling and Limber approximations}).
Hence, we directly model the cross-correlation functions using the expressions outlined below.

\subsection{Angular correlation function}


In photometric surveys we do not have access to accurate galaxy redshifts, and we only have tomographic 2D galaxy maps, $\Delta_g(\nvec)$.
These maps can be described as the integral of the 3D distribution of galaxies along the line of sight, weighted by the redshift distribution of galaxies in the sample:
\begin{equation} 
 \Delta_g(\nvec) = \int \dz ~ n_g(z) \delta_g(\nvec, z) ~+ \Delta_g^{\mu}(\nvec),
 \label{eq:Delta_g1}
\end{equation}
where $n_g(z)$ is the (unknown) normalized distribution of galaxies, and $\delta_g(\nvec, z)$ is the 3D galaxy density contrast \textit{including} the effect of redshift-space distortion (RSD). Although this effect is small for galaxies with a broad redshift distribution, it is significant for the \lyaf. Hence, we model the full RSD effect for consistency.
For the specific expression of $\delta_g(\nvec, z)$ see Appendix~\ref{sec: rsd}. The second term is the modulation of the galaxy density due to lensing magnification of foreground galaxies. Given that the \lyaf is at $z>2$, the lensing efficiency arising from the cross-correlation with galaxies should be very small, hence we ignore this term in the rest of the paper.

Let us define the mean pixel weight as a function of wavelength, averaged over all the quasars within a particular redshift bin $I$:
\begin{equation}
 V_I(\lambda_i) = \frac{1}{\sum_q W^q_I} \sum_{q} W_q(\lambda_i) ~, ~{\forall ~ i \in I}, 
\end{equation}
and therefore, by construction, we have
\begin{equation}
 \sum_{i \in I} V_I(\lambda_i) = 1 ~.
\end{equation}
Following this we define the effective \lyaf redshift distribution in bin $I$
\begin{equation}
 n_F^I(z_i) = \frac{1}{\Delta z^{\rm pix}} V_I(\lambda_i) ~, 
 \label{eq: nfz}
\end{equation}
with $\Delta z^{\rm pix} = \Delta \lambda / \lambda_\alpha = 0.000658$ is the redshift width of the native DESI pixels ($\Delta \lambda = 0.8$ \AA). Example $n_F^I(z)$ with $\Delta z=0.1$ are shown in Fig.~\ref{fig: ng_nFI_SRD_nz_mock_avg} for the raw and \texttt{LyCAN} mocks.


We can estimate the 2D angular cross-correlation between the tomographic galaxy maps and the binned \lya maps with a weighted sum:
\begin{equation}
    \hat{w}_{gI}(\theta) = \frac{\sum_{g,q\in\theta} W^{g} W_{I}^q\Delta_{I}^q}{ \sum_{g,q\in\theta} W^{g} W_{I}^q},
    \label{eq: w_theta}
\end{equation}
where the sum is over all galaxy - quasar pairs with an angular separation $\theta$, and $W^g$ is the galaxy weight. Throughout the paper, we assume the ideal case where $W^g=1$. This gives
\begin{equation}
    \hat w_{gI}(\theta) = \frac{1}{W_{gI}(\theta)} \sum_{q \in \theta} W_I^q \Delta_I^q, 
\end{equation}
where  
\begin{equation}
    W_{gI}(\theta) = \sum_{q \in \theta} W_I^q.
\end{equation}

The expected value of this estimator is
\begin{align}
 \left< \hat w_{gI}(\theta) \right> 
    &= \frac{1}{W_{gI}(\theta)} \sum_{q \in \theta} \sum_{i \in I} W^q(\lambda_i)
        \left< \delta^q_F(\lambda_i) \right> \nonumber \\
    &= \sum_{i \in I} \frac{\sum_{q \in \theta}  W^q(\lambda_i)}{W_{gI}(\theta)} 
        \int \dz^\prime ~ n_g(z^\prime) ~ \xi_{gF}(\theta, z_i, z^\prime) \nonumber \\
    & \approx \int \dz ~ n_F^I(z) \int \dz^\prime ~ n_g(z^\prime) ~ \xi_{gF}(\theta, z, z^\prime)~.
        \label{eq: wgI}
\end{align}

In the second equality we have used Eq.~\ref{eq:Delta_g1}, and $\xi_{gF}$ is the 3D cross-correlation function between the \lyaf and the galaxy sample in redshift space. The expression is given in Appendix~\ref{sec: rsd}.
In the last line, we have used Eq.~\ref{eq: nfz}, assuming that the pixel weights are not correlated with the large-scale structure, and that the background sources distribution is uncorrelated with the galaxies, such that $\sum_{q\in \theta}\sim\sum_q$, and taken the continuous limit on redshift. 




For the convenience of inference on $n_g(z)$, we rewrite Eq.~\ref{eq: wgI} as follows. The 3D cross-correlation function in redshift-space can be written as 
\begin{align}
    \xi_{gF}(\theta, z,z^\prime)=&\int \mathrm{d}^3\mathbf{k} e^{-i\mathbf{k}\cdot\mathbf{r}}
    [b_g(z^\prime)+f(z^\prime)\mu^2]b_F(z)  \nonumber\\
    & \times [1+\beta_F(z)\mu^2]P(\bar{z},k) \nonumber \\
    =&b_g(z^\prime) \int \mathrm{d}^3\mathbf{k} e^{-i\mathbf{k}\cdot\mathbf{r}} b_F(z)[1+\beta_F(z)\mu^2]P(\bar{z},k) \nonumber\\
    & + \int \mathrm{d}^3\mathbf{k} e^{-i\mathbf{k}\cdot\mathbf{r}} f(z^\prime)\mu^2\,b_F(z)[1+\beta_F(z)\mu^2]P(\bar{z},k) \nonumber \\
    =:&b_g(z^\prime)\,\xi_{gF}^b(z^\prime, z, \theta)+\xi_{gF}^{f\mu^2}(z^\prime, z, \theta)~.
\end{align}
We shall assume a cosmology, and that values for the \lya bias and distortion as a function of redshift are known. The only unknown quantity in the equation is the galaxy bias. Hence, we can pre-compute these terms and store these quantities to speed up the computation.
Hence, we can define the following functions:
\begin{equation}
     w_{I}(z^\prime,\theta)=: b_g(z^\prime)\,w_{I}^b(z^\prime,\theta)+ w_{I}^{f\mu^2}(z^\prime,\theta),
     \label{eq: 28}
\end{equation}
where 
\begin{equation}
\begin{aligned}
    w^b_{I}(z^\prime,\theta) &:= \int \dz ~ n_F^I(z) ~\xi^b_{gF}(\theta, z, z^\prime),\\
    w^{f\mu^2}_{I}(z^\prime,\theta) &:= \int \dz ~ n_F^I(z) ~\xi^{f\mu^2}_{gF}(\theta, z, z^\prime).
    \label{eq: wb, wfmu}
\end{aligned}
\end{equation}
And this gives the following form for Eq.~\ref{eq: wgI}:
\begin{align} 
  \langle \hat{w}_{gI} (\theta) \rangle  &= \int \dz^\prime ~ n_g(z^\prime)\,\left( b_g(z^\prime)\, w_{I}^b(z^\prime,\theta)+ w_{I}^{f\mu^2}(z^\prime,\theta)\right).
 \label{eq:bar_w_gI 1}
\end{align}
Notice that the second term contains the RSD effect from the galaxies, and is independent of the galaxy bias.
The second term is small over the range of redshifts, angular scales, and redshift bin sizes considered here, and only reaches about $1\%$ of the first term at the largest $\theta$ bin and at the highest redshifts, where the SNR are small. Hence, in the actual modeling, we essentially ignore the second term, and we treat $n_g(z)b_g(z)$ as fully degenerate. On the other hand, the RSD of \lyaf which is accounted in both terms has a bigger impact. This is not the case for clustering redshifts using galaxies where RSD can be safely ignored \citep{2025A&A...702A.155D}.

In case angular bins are combined through some angular weighting functions (see Section~\ref{sec: angluar avg}), functions similar to Eq.~\ref{eq: wb, wfmu} can be constructed by directly integrating Eq.~\ref{eq: wb, wfmu} with respect to the angles. 

We test and confirm that computing theory with a linear matter power spectrum is sufficient for the scales in this work. We also assume that the \lyaf bias and distortion parameters to be known precisely from the DESI 3D correlation measurements.

We have also neglected lensing magnification effects on the \lyaf. Magnification mainly impacts {the observed flux of the backlight quasar}, which then impacts the \lyaf through the SNR of the observed quasar spectra and the detectability of certain forests. These effects are small on the \lyaf auto- and cross-correlations, hence we drop the relevant term in our model. 

\subsection{Models for \texorpdfstring{$b_g(z)n_g(z)$}{bznz}}

As mentioned above, the cross-correlation function in this case can only constrain the combination $b_g(z)n_g(z)$, as in the low-redshift case. We shall refer to this product as the \textit{bias-weighted redshift distribution}, $bn(z)$.
In this section, we introduce two models to characterize this function: the shift mean model and the Gaussian Process model. 
In both models, we assume a template $bn^T(z)$ for the target function we want to constrain. In conventional methods, this is typically $n(z)$ obtained from photometric calibration methods. Here we use the true $bn(z)$ as the template.

In this paper, we do not implement any methods to separate the bias evolution from the redshift distribution, a problem common to all clustering redshift measurements called `bias mitigation' We discuss implementations with more general models in Appendix~\ref{sec: nz inference general}.

\subsubsection{Shift mean}

The shift mean model is the simplest, and the most widely adopted model in conventional clustering redshift methods. 
The model applies a shift, $\delta z$, to the template such that 
\begin{equation}
    bn(z,\,\delta z)=bn^T(z-\delta z).
\end{equation}
This model aims to calibrate the mean redshift of the distribution, which is most important for weak lensing shear two-point statistics. 

Notice that given our data only cover the tail of the distribution, we do not expect tight constraints on $\delta_z$, and we regard this as the `worst-case scenario'. In a more realistic application, we expect this to be improved by more coverage over the distribution from other reference samples, such as DESI ELGs and QSOs. We shall leave this case for future work.


\subsubsection{Gaussian process (GP)}\label{sec: gp}

A Gaussian process defines a distribution over functions, and it is often used for non-parametric regression problems. A GP is specified by a mean (typically zero) and a kernel function $\mathcal{K}(x,x')$. We adopt the simple radial basis kernel function, where for two points $x,x'$,
\begin{equation}
    \mathcal{K}(x,x'|\sigma_f, \ell)=\sigma_{\mathcal{K}}^2\exp \left( -\frac{(x-x')^2}{2\ell^2} \right ),
\end{equation}
and the $\sigma_{\mathcal{K}}$ and $\ell$ parameters control the amplitude and length scales at which the GP models vary. 

In this case, we apply a GP to characterize the fractional variation around the template with $N_{\rm GP}=3$ nodes. These nodes are points in redshifts, $\mathbf{z^{\rm node}}$, varying only within the range $[z_{\rm min}, z_{\rm max}]$ where we have the \lyaf data:
\begin{equation}
    bn(z, \{\mathbf{x} \}) = bn^T(z) \exp(\mathcal{GP}[\mathcal{K}(z, \mathbf{x})]),
    \label{eq: gp}
\end{equation}
where $\mathbf{x}$ is a vector containing GP latent variables of dimension $N_{\rm GP}$, drawn from a standard normal distribution. We tested increasing the node number and it does not change the posterior on the variation of $bn(z)$. We take the exponential in Eq.~\ref{eq: gp} to avoid negative values in $bn(z)$. Notice that this results in a very skewed prior towards larger values compared to $bn^T(z)$. The GP can be written as 
\begin{equation}
    \mathcal{GP}[\mathcal{K}(z, \mathbf{x})] = (\mathbf{K}^T \mathbf{L}^{-1})\mathbf{x},
\end{equation}
where $\mathbf{K}$ is a kernel matrix $K_{ij} = \mathcal{K}(z^{\rm node}_{i},z_j|\sigma_{\mathcal{K}}, \ell)$, and $\mathbf{L}$ is the Cholesky decomposition of the kernel matrix at the nodes, $K^{\rm node}_{ij} = \mathcal{K}(z_i^{\rm node}, z_j^{\rm node}|1, \ell)$. 

The prior of this model depends on the parameters $\sigma_{\mathcal{K}}$ and $\ell$. Hence, in typical GP inference problems, one needs to optimize or marginalize over these parameters. Here, given that the tail of our target $bn(z)$ is featureless, we test a few values of $\ell$.
The GP results can also depend on the kernel selection. We do not test other kernels due to the simplicity of the target function used here. We set $\sigma_{\mathcal{K}}=1$, $\ell=0.01,0.05,0.1$, and $\mathbf{z^{\rm node}}=(2.4,2.6,2.8)$. Notice that a large $\ell$, meaning a large correlation length, also extends the prior uncertainty range into regimes that are not covered by the measurements (e.g. with $\ell=0.1$, the GP model can vary from $z\sim 1.5$, which is hardly informed by the \lya data). Hence if one wishes to put a strict boundary to the prior, one would also need to tune $\mathbf{z^{\rm node}}$. We do not do this tuning here, and we leave detailed investigation of the GP model in more realistic studies for the future.

This setup represents the `best-case scenario' for the constraints on $bn(z)$, where the function outside of the data coverage is given by the truth and is known exactly.

\section{Measurements}
\label{sec: estimators}

\subsection{Estimators with mean subtraction}
\label{sec: estimators detail}




The angular cross-correlation function can be measured using the estimator in Eq.~\ref{eq: w_theta}. However, this estimator can be impacted by the mean value of the flux transmission fluctuation, $\langle \Delta_I^q\rangle$, over the footprint in each redshift bin $I$, which is of similar order of magnitude as $w_{gI}$. 
Typically, $\delta_F^q(\lambda)$ has a zero mean over $\lambda$ by construction when averaged over the DESI footprint. However, this is not true for the joint DESI$\times$LSST footprint. The mean value $\langle \Delta_I^q\rangle$ leads to an additive bias that dominates the cross-correlation signal, hence needs to be subtracted in the estimator\footnote{This is also true for Jackknife regions. Estimators without the mean subtraction give rise to significantly larger error bars and a more correlated covariance matrix (see Appendix~\ref{sec: randoms}).}.


One way of dealing with this is to swap the $\Delta_{I}^q$ in Eq.~\ref{eq: w_theta} for
\begin{equation}
    \tilde{\Delta}_{I}^q = \Delta_{I}^q - \frac{\sum_{q \in {\rm mask}}  W^q_I \Delta_{I}^q}{\sum_{q \in {\rm mask}}  W^q_I },
    \label{eq: w_theta2}
\end{equation}
where the sum is over all lines of sight within the sub-mask.

Another way is to include a uniform random catalogue over the sub-mask, and modify the estimator as 
\begin{equation}
    \tilde{w}_{gI}(\theta) = w_{gI}(\theta) - \frac{\sum_{R,q\in\theta} W_{I}^q\Delta_{I}^q}{ \sum_{R,q\in\theta} W_{I}^q},
    \label{eq: rand subtraction}
\end{equation}
where the sum is over all random - quasar pairs separated by $\theta$. 
We test and compare these estimators in Appendix~\ref{sec: randoms}, and we use Eq.~\ref{eq: rand subtraction} as the fiducial estimator in this work. This is implemented in the python package \texttt{yet\_another\_wizz}\footnote{\url{https://github.com/jlvdb/yet_another_wizz}}, which we use to measure the cross-correlation function.





Notice that in all of our cross-correlation estimators, the denominator in Eq.~\ref{eq: w_theta} essentially counts galaxy-quasar pairs, which themselves cluster. In similar galaxy-field cross-correlation studies, such as weak lensing, the random catalogue is also used to estimate this denominator term to avoid biasing the correlation function, an effect referred to as `lens-source clustering'. Given that this is a small secondary effect, we do not introduce any changes to our estimators here. 
The study of the best estimator for \lya-galaxy cross-correlation function is beyond the scope of this paper, and we leave this for future work.

\subsection{Angular average}
\label{sec: angluar avg}


We adopt $15$ logarithmic $\theta$ bins with range $\theta=[1,50]$ arcmin. At the \lyaf reference redshift, $z=2.4$, this corresponds to a comoving scale of $\chi=[1.15, 57.53]h^{-1}$Mpc.

In conventional clustering redshift methods, angular bins are combined via a scale-dependent weighting function $W_{\rm scale}(\theta)$:
\begin{equation}
    \bar{w}_{gI} = \int_{\theta_{\rm min}}^{\theta_{\rm max}} W_{\rm scale}(\theta)\,\tilde{w}_{gI} (\theta) \, \mathrm{d}\theta.
    \label{eq: ang comb}
\end{equation}
The angular weight is typically a power-law, $W_{\rm scale}({\theta}) \propto \theta^{\alpha}$, with $\alpha \sim -1$ adopted to enhance the signal-to-noise from small scales. An alternative choice is to measure the correlation function at a fixed comoving distance $r$, which is converted from the angular scale $\theta$ assuming a fixed cosmological model. We do not adopt this approach here because we would like to keep the measurements independent of cosmology. 
In this paper, we additionally explore the option to fit multiple scales simultaneously.

We compute the model at the angular scales that correspond to the mean of the $\theta$ bins, and if needed, we use Eq.~\ref{eq: ang comb} to combine the scales, and compare with the measurements. 
Given that we adopt logarithmic bins, the bin size increases with $\theta$. We sub-divide the larger angular bins and test that computing the models at the mean $\theta$ in each bin is accurate enough for the purpose of this paper.

\subsection{Covariance and likelihood}

We obtain the error bars and the covariance matrix of the measurements by Jackknife sampling. The footprint is divided into $N_{J}$ regions of equal area, and the measurements are repeated multiple times, each with one region left out. 
Given the data vector $\mathbf{d}^k$ for a Jackknife sample $k$, the covariance matrix $\mathcal{C}$ is estimated by
\begin{equation}
    \mathcal{C}_{ij}=\frac{N_{J}}{N_{J}-1} \sum_{k=1}^{N_J} [d^k_i-\bar{d}_i] [d^k_j-\bar{d}_j],
\end{equation}
where $i,j$ are elements of the covariance and the data vector, and $\bar{\mathbf{d}}$ is the average of the Jackknife samples.
We have taken $N_{J}=64$ in the main analysis, and we tested that changing it to $N_J=128$ does not significantly impact the covariance.

\begin{figure*}
    \centering
    \includegraphics[width=\linewidth]{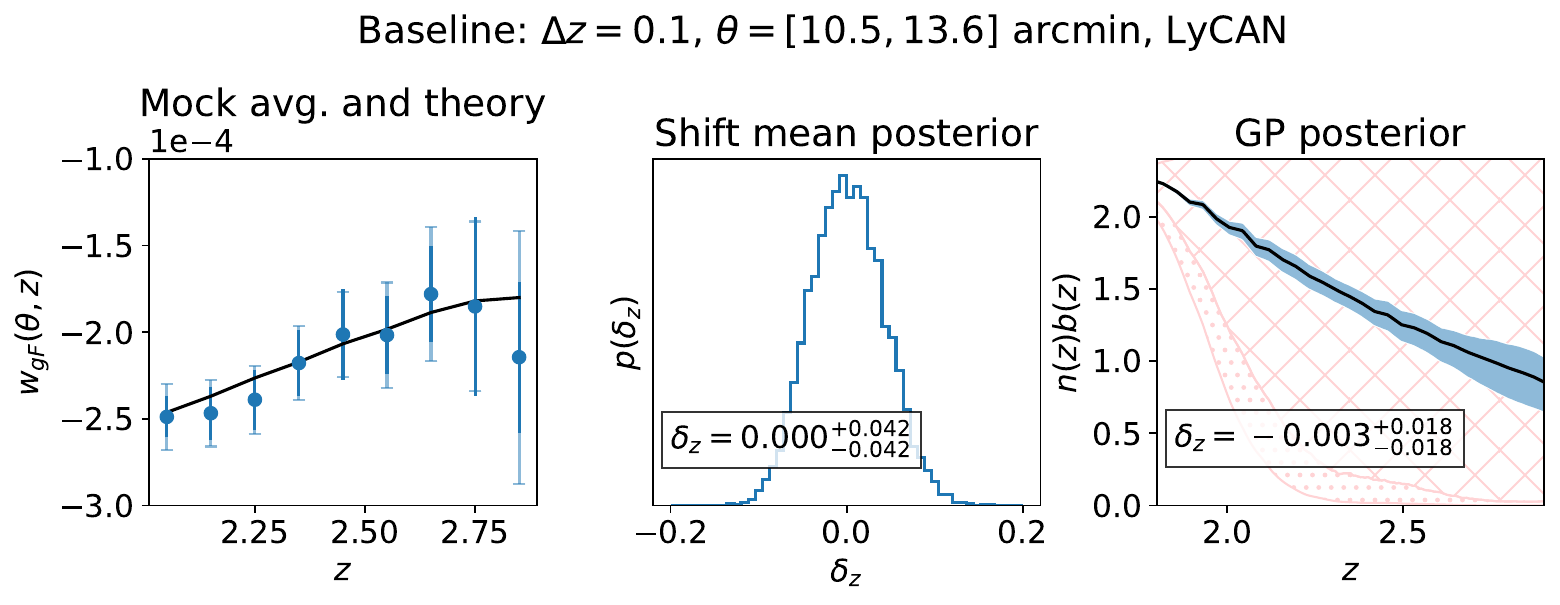}
    \caption{Results of the baseline setup: we use the \texttt{LyCAN} mocks, adopt a redshift bin size of $\Delta z=0.1$, and measure the angular cross-correlation function on the scale $\theta=[10.5, 13.6]$ arcmin. \textit{Left panel}: the data vector from the stack of 10 \texttt{CoLoRe} mocks (blue data points). The solid error bars correspond to the standard deviation of the 10 mocks, while the faint error bars correspond to the average of the Jackknife covariance. Both of these error bar estimates are quite noisy.
    The theory prediction is shown as the black solid line. \textit{Middle panel}: the posterior of the shift parameter $\delta_z$ in the shift mean model using synthetic data vector. The best-fit value and the 68\% confidence interval is shown in the box. \textit{Right panel}: the 1$\sigma$ contour (blue shaded region) on $b(z)n(z)$ from 5000 posterior samples of the Gaussian Process (GP) model. The true $b(z)n(z)$ is shown by the black solid line. The pink shaded regions with cross and dots mark the 68\% and 95\% prior range, where the GP model is allowed to vary. The constraints on the mean redshift $\delta_z$ is shown in the box.}
    \label{fig: baseline_results}
\end{figure*}

Given the model $w_{gI}$ with a set of parameters $\mathbf{s}$, the measurement $\hat{w}_{gI}$, and the covariance $\mathcal{C}$, we can evaluate the likelihood: 
\begin{equation}
    \mathcal{L}(\{\hat{w}_{gI}\}\, \vert \, bn^T, \mathbf{s})\propto\exp\left[-\frac{1}{2}\left( \hat{w}_{gI} - w_{gI} \right)^\top \mathcal{C}^{-1}\left(\hat{w}_{gI}- w_{gI} \right)\right].
\end{equation}
With Bayes' theorem, we can evaluate the posterior probability of the model parameters 
\begin{equation}
    p(\mathbf{s} \,\vert \{\hat{w}_{gI}\},\, bn^T)\propto \mathcal{L}(\{\hat{w}_{gI}\}\, \vert \, bn^T,\mathbf{s}  )\,\pi(\mathbf{s}), \label{eq:Ldzs}
\end{equation}
where $\pi(\mathbf{s})$ is the prior for the model parameters. For the shift mean model, we take a uniform prior over $\delta_z=[-0.5,0.5]$, and for the GP model, we use the parameters listed in Section~\ref{sec: gp}. The prior range on $bn(z)$ can be found in Fig.~\ref{fig: baseline_results}.

Given that we have 10 mocks, we also estimate the variance of the data vector via the scatter of these 10 mocks, and compare with the Jackknife errors.
We show both sets of error bars in our results. Finally, for the covariance matrix, we use the average Jackknife covariance between 10 mocks to reduce noise. 

{Finally, we use the synthetic data vector, i.e. the data vector computed from theory, instead of the actual mock measurements, for the inference step.} This is because we observe some unexpected fluctuations in the mock data vector as a function of redshift (see Appendix~\ref{sec: validation results} for more details).

\section{results}
\label{sec: results}

We first present the baseline results in Section~\ref{sec: validation}, showing a relatively realistic and optimal case. In the subsequent sections, we vary each analysis choice in the baseline results. Particularly, in Section~\ref{sec: Noise, continuum fitting, and contamination} we show how the data vector and constraints on $bn(z)$ changes with different \lyaf mocks, i.e. the raw, true continuum, \texttt{Picca}, \texttt{LyCAN}, and contaminated \texttt{LyCAN} mocks. In Section~\ref{sec: Angular scales}, we investigate the impact of angular scales of the cross-correlation measurements. In Section~\ref{sec: Redshift bin size of the forest}, we vary the redshift bin size $\Delta z$ of the cross-correlation measurements. 

Because both tracer densities drop significantly beyond $z=2.9$ in the simulation, we remove any redshift bins above $z=2.9$. This results in 9 redshift bins with $\Delta z=0.1$, and 18 redshift bins with $\Delta z=0.05$.

\subsection{Baseline results}
\label{sec: validation}

We present our main result of the paper in Fig.~\ref{fig: baseline_results}, which represents a realistic forecast for using DESI Y5 \lyaf{s} to calibrate the redshift distribution of the tomographic bin of the LSST Y10 highest-redshift source sample. 
In this setup, we use \texttt{LyCAN} mocks and adopt a redshift bin size of $\Delta z =0.1$ between $z=[2,3]$, providing reasonable resolution for the high-redshift tail of the galaxy redshift distribution. We use the \lyaf { } -- galaxy cross-correlation measurement at the smallest usable angular bin, at $\theta=[10.5, 13.6]$ arcmin, which provides nearly optimal signal-to-noise. 

We obtain a $24\sigma$ detection of the clustering redshift signal in this baseline setup.
The left panel of Fig.~\ref{fig: baseline_results} shows the measured data vector from the average of 10 mock realizations. We show two sets of error bars: the standard deviation of the 10 mocks as solid vertical lines, and the average Jackknife error bars in faint lines with caps. The Jackknife error bars are consistently larger than the sample variance from the mocks. The error bars shown here are equivalent to one realization. The theory expectation is shown by the solid black line.

We show the mean shift model constraints on $\delta_z$ using synthetic data and the average Jackknife covariance matrix in the middle panel of Fig.~\ref{fig: baseline_results}. We can constrain this parameter to $\delta_z=0.000\pm0.042$, equivalent to an error of $\sigma_z/(1+\bar{z}) = 0.014$ (taking the mean redshift of the bin to be $\bar{z}=2$). This is a magnitude larger than the Y10 LSST DESC requirement, $\sigma_z/(1+\bar{z}) = 0.001$, but we stress that this is only using data provided at $z=[2,3]$. In practice, the lower redshift sample will be calibrated by clustering redshifts with other spectroscopic tracers. Altogether the shift model should be constrained to a much better precision. We leave this case for future investigation. 

The right panel of Fig.~\ref{fig: baseline_results} shows the constraints using the Gaussian Process (GP) model, with $\ell=0.05$. We sample from the posterior on the three GP nuisance parameters to obtain the posterior on $bn(z)$. The 68\% contours are shown in light blue.
We also map the prior onto $bn(z)$, and show its 68\% and 95\% contours in pink shades. 
For each posterior sample, we compute the shift in the mean redshift to produce an equivalent constraint on the shift parameter, $\delta_z$. We obtain a constraint of $\delta_z=-0.003\pm0.017$. This corresponds to $\sigma_z/(1+\bar{z})=0.006$, closer to the LSST DESC requirement. Changing the $\ell$ values also introduces small changes to the posterior on $\delta_z$, at a level of $5-8\%$ given the values we test here. However, these changes mainly come from the low-redshift part without data coverage due to the prior, rather than the constraining power from the data. 
This also shows how much the uncertainty in the high-redshift tail can shift the mean redshift of the tomographic bin, easily exceeding the required error budget. 

\begin{figure*}
    \centering
    \includegraphics[width=0.9\linewidth]{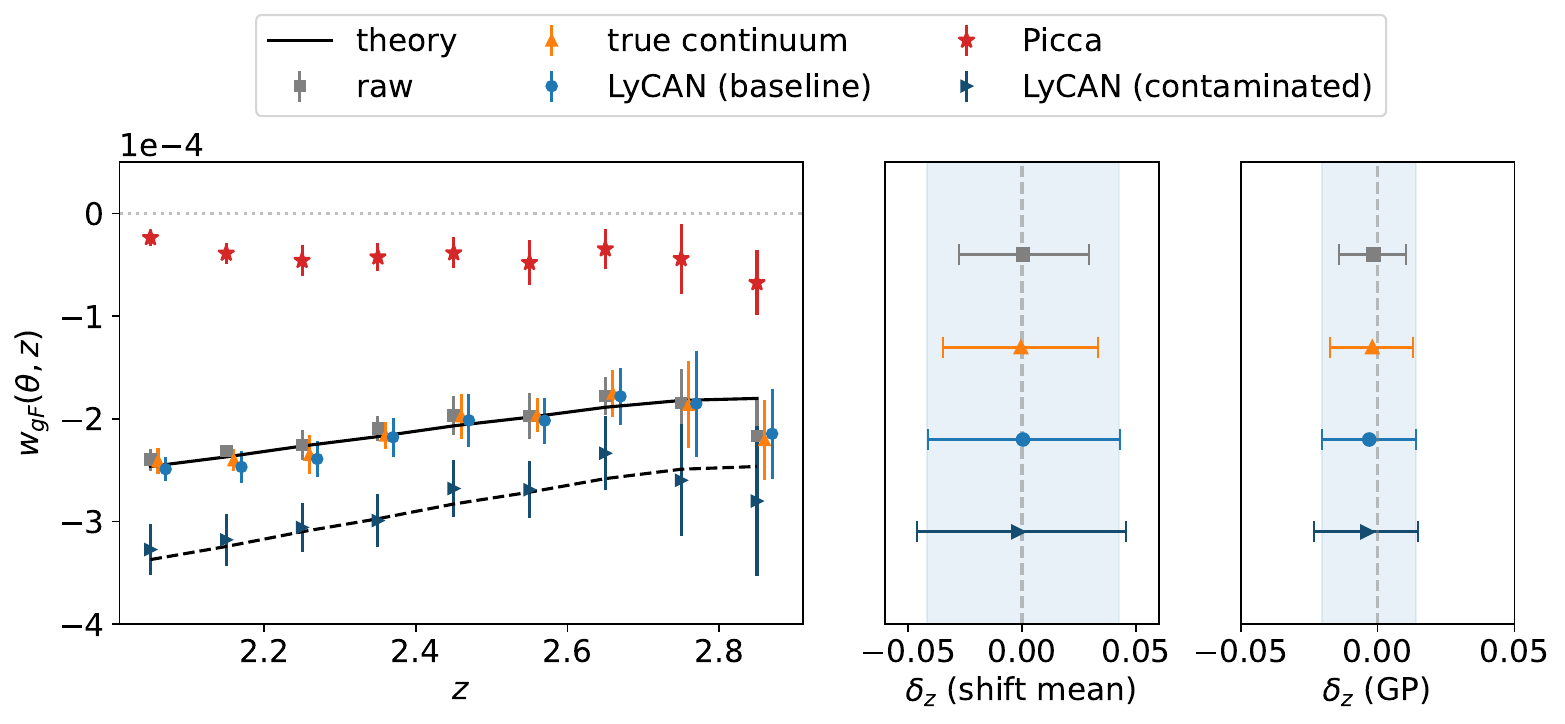}
    \caption{\textit{Left panel}: The \lya-galaxy cross-correlations measured from 10 \texttt{CoLoRe} mocks with different forest implementations: the noiseless truth (`raw', grey squares), the noisy case with known continuum (`true continuum', orange triangles), the noisy case with \texttt{LyCAN} continuum subtraction (`LyCAN (baseline)', blue circles), the noisy case with \texttt{Picca} continuum subtraction (`Picca', red stars), and the contaminated case with \texttt{LyCAN} continuum subtraction (`LyCAN (contaminated)', navy triangles). The theory prediction is shown as the black solid line. For the contaminated case, the data points can be described with the same theory, with an increased effective bias by 25\%, shown by the black dashed line. \textit{Middle panel}: increase in the error bars on the shift mean parameter as more realism is added. \textit{Right panel}: increase in the constraints on the mean redshift derived from the Gaussian Process (GP) model posterior.}
    \label{fig: delta_modes}
\end{figure*}

\subsection{Noise, continuum fitting, and contamination}
\label{sec: Noise, continuum fitting, and contamination}

In this section, we use the various mocks described in Sections~\ref{sec:simulations} \&~\ref{sec: Estimating the fluctuations} to assess the impact of noise, continuum fitting methods, and contamination of the \lyaf on the cross-correlation function, as well as on the $\delta_z$ constraints. 

Figure~\ref{fig: delta_modes} shows the results of the test. The left panel shows the cross-correlation measurements for various \lyaf cases. We fix the redshift binning and angular scales to the same as the baseline. {The data points shown are the average and the scatter between the 10 mocks.} The raw, true continuum, and the \texttt{LyCAN} mocks show good agreement with the theory expectation with increasing error bars. The true continuum mocks yield a $30\sigma$ measurement, corresponding to the best case we could achieve given DESI-like noise.
On the other hand, the \texttt{Picca} continuum fitting introduces significant distortion to the shape of the cross-correlation function, and the signal-to-noise (SNR) drops to $14\sigma$. 
Although the detection is still statistically significant, modeling of this case would require a more complicated model involving the distortion matrix for \texttt{Picca} to account for the lost large-scale signal.
This highlights the importance of \lya continuum fitting for clustering redshifts. 

Finally, we show the case with contamination included in the spectrum. Most contamination identified in the spectra are masked, but there can still be residuals or unidentified contamination. We follow \cite{DESI:2025qqu} to mask BAL and DLA contamination, and to model the remaining contaminants, including metal absorption and redshift errors. The contaminants introduce another population of biased tracers to the \lyaf, effectively changing the \lyaf bias \citep{2012JCAP...07..028F}. In this case, the contaminated measurement has a $25\%$ larger overall bias compared to the fiducial bias. The SNR of the measurement is $22\sigma$, only slightly degraded from the baseline.

The middle and right panels of Fig.~\ref{fig: delta_modes} show the degradation in the constraints of the shift $\delta_z$ in the mean redshift of the tomographic bin for the shift mean model and the GP model, respectively. We see a very consistent trend. The contaminated case results in a $\sim5\%$ error increase compared to the baseline, assuming the effective bias of the contaminated \lyaf is known.

\subsection{Optimal angular scales}
\label{sec: Angular scales}

\begin{figure}
    \centering
    \includegraphics[width=\columnwidth]{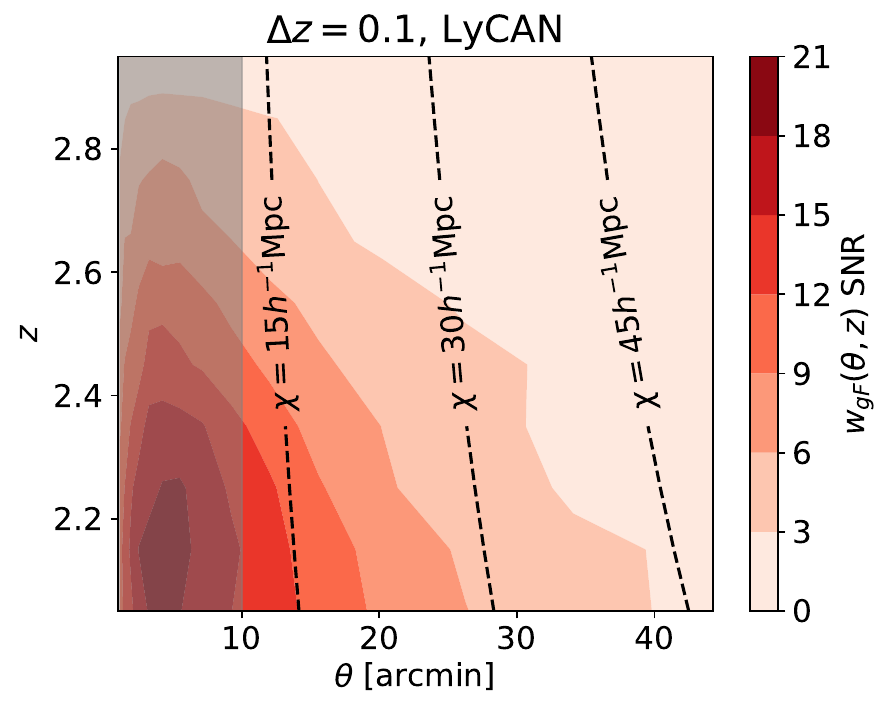}
    \caption{Signal-to-noise ratio (SNR) of the \lya-galaxy angular cross-correlation for the \texttt{LyCAN} continuum fitting mocks, as a function of redshift and angular scale. The shaded region marks the minimum scales we do not use, below which the simulations deviate significantly from the theory due to lack of power at small scales. The three dashed lines show constant comoving scales of $15,30,45h^{-1}$Mpc for reference.}
    \label{fig: LyCAN_snr}
\end{figure}

\begin{figure}
    \centering
    \includegraphics[width=\columnwidth]{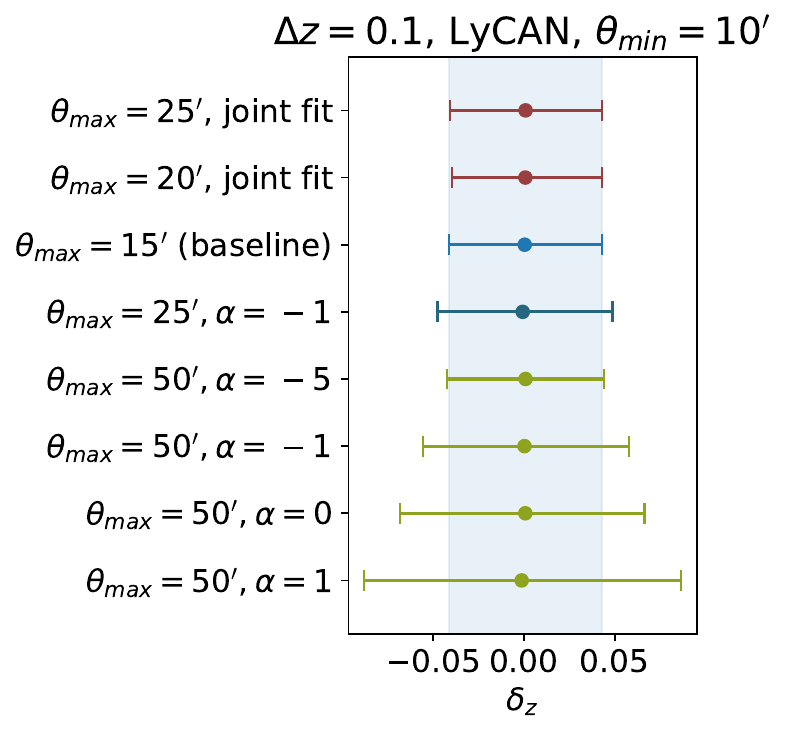}
    \caption{The impact of different angular scales of the cross-correlation function on the mean shift constraints on the unknown galaxy $n(z)$. The lower four points (green) use all measured scales between $\theta=[10, 50]$ arcmin. The compressed data point per redshift bin is the weighted average of each angle, with a weight $\propto \theta^{\alpha}$. The dark blue point denotes a similar case, but with $\theta_{\rm max}=25$ arcmin. The baseline is shown in light blue, where the single data points with $\theta_{\rm max}=15$ arcmin is used. The two red points show the case where we jointly fit two and three angular bins (denoted by $\theta_{\rm max}=20, 25$ arcmin, respectively). }
    \label{fig: theta_scales}
\end{figure}

In this section we test different angular scale combinations to find a configuration with optimal signal-to-noise in terms of the constraints. All other analysis choices are the same as the baseline. 

Before we explore further the optimal angular combination, we show the SNR of the measured cross-correlation function as a function both redshift and angular scales in Fig.~\ref{fig: LyCAN_snr}. The SNR decreases with redshift as both tracer samples decrease in number density. The SNR increases as we go to smaller angles, and reaches a maximum at $\theta \sim 5$ arcmin. However, we only use scales larger than 10 arcmin. This cut is set by the limitations of our simulations, where the correlation function starts to become lower than the linear theory prediction due to the smoothing of the simulation at small scales (see Appendix~\ref{sec: validation results}). 
Hence, we stress that this work only provides an initial feasibility test for this methodology given these limitations, and the measurement could potentially be improved by taking smaller scales than used here. However, it should also be investigated, with a more realistic mock catalogue, when the linear modeling breaks down, and whether the knowledge of the linear \lyaf forest bias and distortion extrapolates to these scales, as they are measured at much larger scales (typically $50 - 100h^{-1}$Mpc).
We leave this for future work.  

Now we turn to the task of finding optimal angular combinations. As mentioned in Section~\ref{sec: angluar avg}, in conventional clustering redshift methods, different angular scales are typically combined via a scale-dependent power-law weight with power $\alpha$. $\alpha<0$ up-weights the small scales, and vice versa. Here we test the following scenarios, in addition to the baseline case:
\begin{enumerate}
    \item Use all angular bins between $\theta_{\rm min}=10$ arcmin, and $\theta_{\rm max}=50$ arcmin, and vary $\alpha=-5, -1,0,1$;
    \item Fix $\alpha=-1$, set smaller $\theta_{\rm max}=20, 25$ arcmin;
    \item Fit multiple angular bins jointly by concatenating the data vectors. We vary $\theta_{\rm max}=20, 25$ arcmin, which includes 2 and 3 angular bins respectively. We include the correlation between different angular scales in the covariance.
\end{enumerate}

The results are shown in Fig.~\ref{fig: theta_scales}, and we use the constraints on $\delta_z$ for the shift mean model as an example. Scenario (1) is shown by the set of green points. The constraint gets tighter as $\alpha$ becomes more negative. This is consistent with Fig.~\ref{fig: LyCAN_snr} where smaller scales have higher SNR. We specifically included a particularly small value, $\alpha=-5$, to see where the uncertainty on $\delta_z$ converges. With this scaling, nearly all of the weight is given to the very first angular bin, and it clearly gives a smaller constraint compared to putting more weight on the larger scales. To avoid repetition, we only show the case where $\theta_{\rm max}=25$ arcmin for scenario (2) by the dark blue point. We see that compared to the case with $\theta_{\rm max}=50$ arcmin with $\alpha=-1$, cutting off large scales improves the constraints. However, it is not as competitive as the single bin case, shown by the light blue point. This is counter-intuitive because one would expect the constraint to improve with more information added from larger scales. This is possible when the multi-scale information in the correlation function is not optimally compressed. 

To investigate whether we can further improve the constraints by leveraging the multi-scale measurement, we go to scenario (3), which is shown by the dark red points. The improvement in the constraints in this scenario is marginal compared to the baseline. This may be explained by the covariance matrix of the concatenated data vector, where the correlation between different angular scales within the same redshift bin can be as high as $80\%$ for neighbouring bins, and $70\%$ for the next-nearest neighbour.

We conclude that the optimal scale for clustering redshift measurements is the smallest angular bin. We note that with an improved simulation, one may use smaller scales than $\theta_{\rm min}=10$ arcmin to obtain tighter constraints.

\subsection{Redshift bin size of the forest}
\label{sec: Redshift bin size of the forest}

\begin{figure*}
    \centering
    \includegraphics[width=\linewidth]{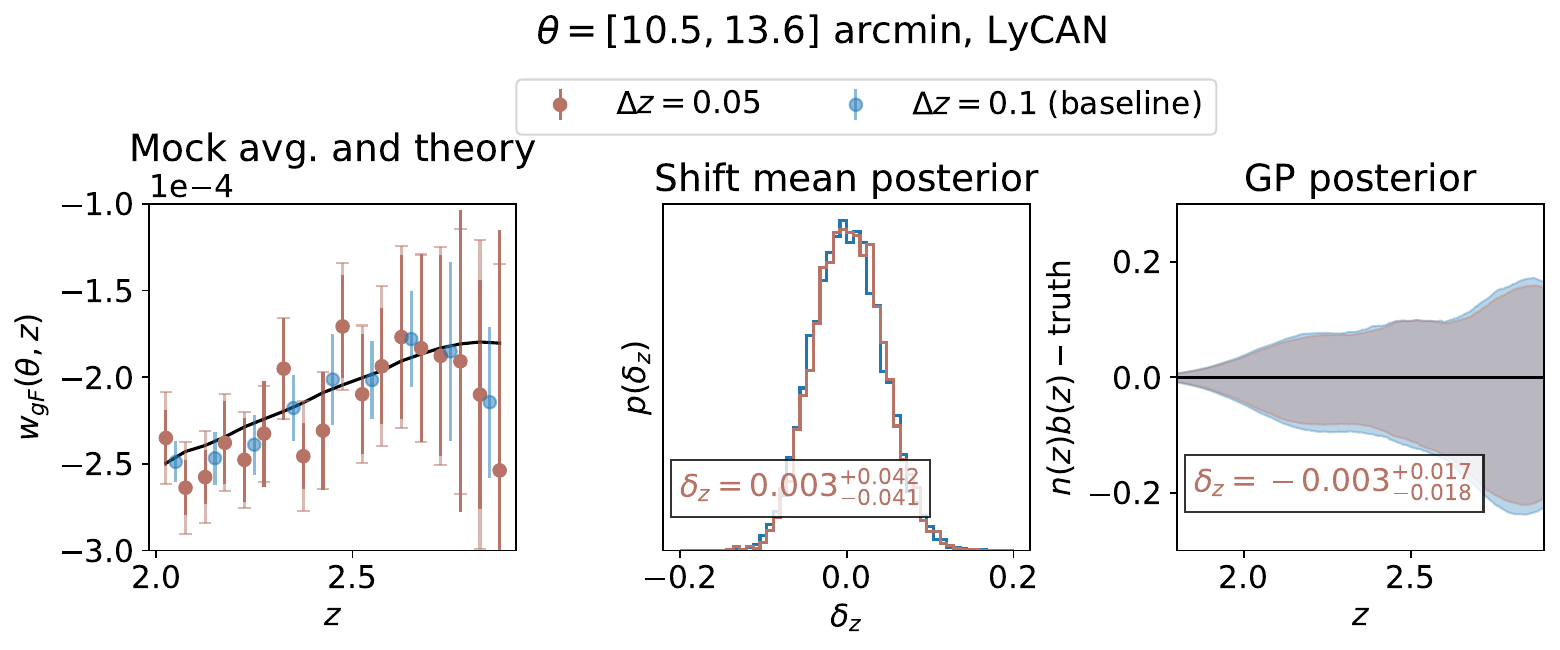}
    \caption{The impact of changing the redshift binning. The $\Delta z=0.05$ and baseline $\Delta z=0.1$ cases are denoted by red and blue, respectively. The solid and faint error bars in red denote scatter from 10 mocks and the average Jackknife error for $\Delta z=0.05$ case, whereas only the error bar from mock scatter is plotted for $\Delta z =0.1$ for readability. \textit{Left panel}: measured cross-correlation function at $\theta=[10.5,13.5]$ arcmin for the \texttt{LyCAN} continuum fitting. \textit{Middle panel}: constraints on the mean shift parameter. \textit{Right panel}: the difference between the true $n(z)$ and those sampled from the posterior of the GP model with three free parameters. They grey region is a result of the overlap between the two posteriors. Both constraints are not impacted by the bin size.}
    \label{fig: zbin_compare}
\end{figure*}

In this section, we vary the redshift bin size of the \lyaf. There are pros and cons to having large and small redshift bin sizes. 
A large redshift bin size effectively reduces the impact of RSD in the \lyaf (mainly in the auto-correlation, see Fig.~\ref{fig: rsd effect}). 
If one also chooses to measure at very small scales, then one may be in a regime where the conventional clustering redshift estimator with the auto- and cross-correlation ratios is accurate enough. 
Moreover, most high-redshift tails are relatively featureless. A small number of redshift bins is sufficient to constrain the shape of the distribution well. 
On the other hand, a small redshift bin size allows one to sample the shape of the galaxy redshift distribution with more resolution, which may be useful for certain galaxy samples. The advantage of the \lyaf in this case is that the SNR in each bin is not significantly degraded as a cost of slicing the reference sample finely in redshift. This is the case, however, for galaxy reference samples, such as quasars, where halving the redshift bin size means reduces the number density and increases the shot noise of the correlation function. 
Notice that the choice of bin size does not concern our modeling because we do not use the Limber approximation.

In Fig.~\ref{fig: zbin_compare}, we show the measurement with a redshift bin size of $\Delta z=0.05$, i.e., half of the baseline set up. We keep all other analysis choices the same as the baseline. The left panel shows the measured data vector. We show two sets of error bars: one from the scatter of 10 mocks, and the other from the average Jackknife error. The SNR of the cross-correlation measurement is $23\sigma$, very close to the baseline. 
The theory expectation is shown in black. We also include the baseline case for comparison. 
In the middle panel, we compare the shift mean posteriors for both cases, and these are very consistent. The GP model with $\ell=0.05$ tells a similar story on the right panel. Here we show the difference between the 68\% posterior samples of the modeled $b(z)n(z)$ and the truth, to make the difference between the two cases more visible. With a smaller correlation length, $\ell=0.01$, the constraint gets tighter by about $8\%$, again due to the reduction of uncertainty below and near $z=2$. At $z>2$, the posterior on $bn(z)$ is consistent in both cases.

This test shows that given the current setup, a bin size of $\Delta z=0.1$ is sufficient to map out the tail of the unknown biased redshift distribution $bn(z)$. 
This is because the $bn(z)$ tail is featureless such that the neighboring bins are highly correlated, containing very similar values. Hence, we do not lose any information by combining them by adopting a larger binning scheme.

\subsection{SNR as a function of survey area and tracer number density}

\begin{figure}
    \centering
    \includegraphics[width=\columnwidth]{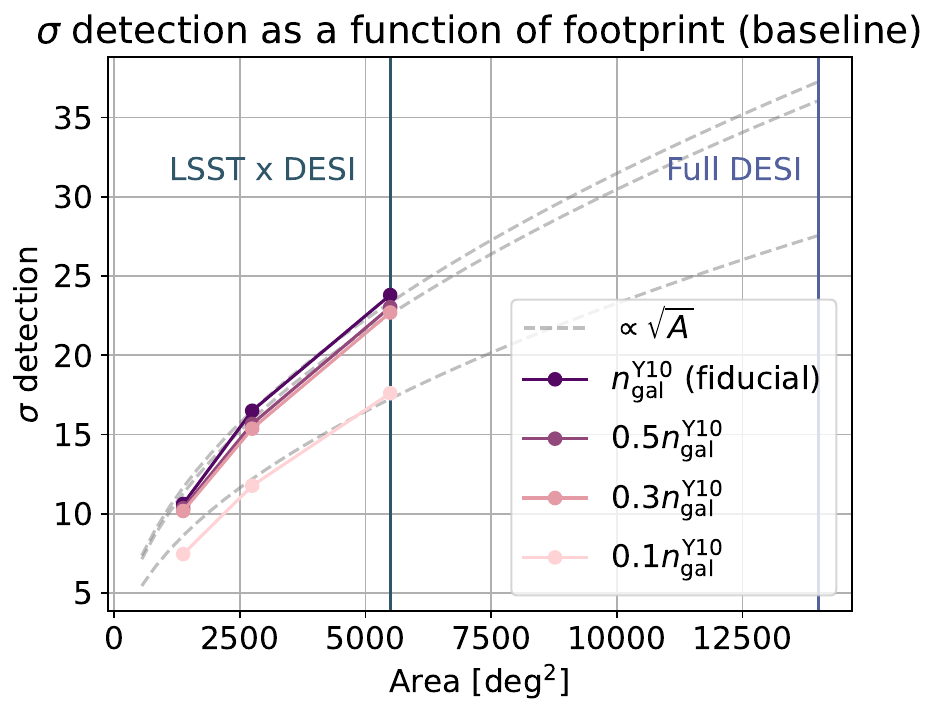}
    \caption{The detection significance of the \lya - galaxy cross-correlation in the baseline setup as a function of the survey area and the unknown galaxy sample number density. The fiducial number density of $n_{\rm gal}^{\rm Y10}=5.4\,{\rm arcmin}^{-2}$ is shown by the dark purple dots, while the purple and pink dots show cases where the galaxy number density is $0.5$, $0.3$ and $0.1$ times the fiducial case, respectively. The grey dashed curves show a fitted $\sqrt{A}$ relation between the SNR and the survey area $A$. The vertical blue lines mark the LSST$\times$DESI survey area and the full DESI area.}
    \label{fig: area_ngal}
\end{figure}

\begin{figure}
    \centering
    \includegraphics[width=\columnwidth]{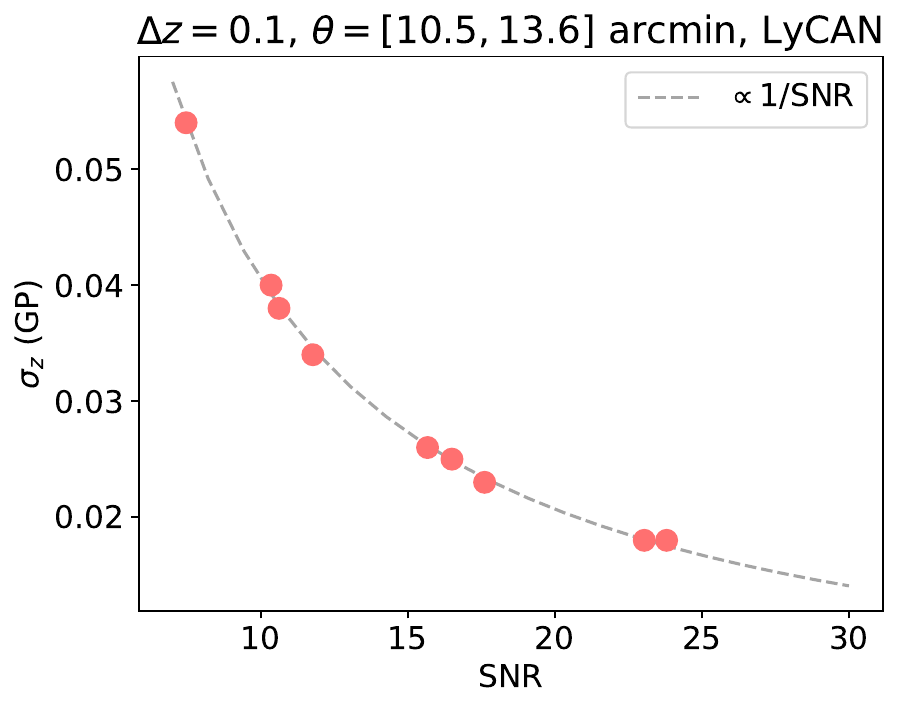}
    \caption{{The uncertainty on the mean redshift, $\sigma_z$, as a function of the signal-to-noise ratio (SNR) of the cross-correlation measurements, obtained with the Gaussian Process (GP) model. The data points are taken from varying survey area and unknown galaxy number density as in Fig.~\ref{fig: area_ngal}. The dashed line shows a fit that scales with $1/{\rm SNR}$. The baseline corresponds to the rightmost point.}}
    \label{fig: snr_dz_relation}
\end{figure}

In this section we investigate how the signal-to-noise (SNR) of the baseline measurements changes with survey area and the unknown galaxy number density. We hope to provide a rough guide for generalizing the results of this paper to other surveys, or earlier LSST data, with different area and galaxy number density. Notice however, that in this exercise we do not change the unknown galaxy redshift distribution. For earlier LSST data, for example, the fraction of objects and the overall number density at $z>2$ will be lower, reducing the signal (not noise) from the one shown here.
We also do not change the \lyaf number density nor its noise level. For an earlier DESI data release, we expect the quasar spectra to be measured with a lower SNR, hence degrading the flux transmission field used in these analysis, reducing the SNR. 

Figure~\ref{fig: area_ngal} shows how the SNR of our baseline setup changes with survey area and the unknown galaxy number density. We measure the SNR for the cases where the footprint is $0.5$ and $0.25$ of the fiducial LSST$\times$DESI footprint. We find that both the diagonal terms of the covariance matrix and the SNR scale well with $\sqrt{A}$, where $A$ is the survey area. This is consistent with what we expect from similar studies using spectroscopic galaxies, where the SNR scales as the number of spectroscopic galaxies, $N_{\rm spec}$ \citep[e.g.,][]{2013MNRAS.433.2857M, 2025A&A...702A.155D}.
With $0.25$ of the fiducial area, corresponding to $\sim 1800 \,{\rm deg}^2$, the baseline setup gives a $10\sigma$ detection.
Extrapolating this relation gives a $37\sigma$ detection over the full DESI footprint in the baseline setup.

We also subsample the unknown galaxy to $0.5$, $0.3$ and $0.1$ times the fiducial number density and measure the detection significance. We find that taking half or one-third of the number density does not significantly reduce the SNR of the measurement, meaning that the uncertainty is dominated by the \lyaf noise, rather than galaxy shot noise. In this regime, the SNR is related to $n_g$ via a linear relation $\sigma = s( n_g / n_{\rm gal}^{\rm Y10})+\sigma_0$, where the slopes and the constants for $A/A^{\rm baseline}=0.25, 0.5, 1$ are measured to be $(s, \sigma_0) = (0.6, 10), (1.6, 15), (1.6, 22)$, respectively. Reducing the number density to $0.1$ times the fiducial, i.e. $n_{\rm gal}=0.54 \, {\rm arcmin}^{-2}$, results in a sharp drop in detection significance (e.g. $17.5\sigma$ over the full footprint). Notice that this is comparable to the number density of Lyman-break Galaxies (LBG) through $u$-band dropout. For LSST Y10, the expected number density is $n_{\rm gal}=0.89 \, {\rm arcmin}^{-2}$ \citep{2025arXiv250306016C}. However, notice that LBGs have a very different $n_{\rm gal}(z)$, which consists of a small population at very low redshift, and the main population peaking at $z=3$ with a width of $\Delta z\sim0.3$, and the galaxy bias is expected to be higher than magnitude limited samples \citep{2019JCAP...10..015W, 2024JCAP...08..059R, 2025JCAP...05..031P}.
Hence, the SNR for the actual application to LBGs may differ from the results here. We leave detailed investigation of this for future work.
{In Fig.~\ref{fig: snr_dz_relation}, we relate the SNR of the above measurements to the uncertainty on the mean redshift, $\sigma_z$. Given that the uncertainty of the mean redshift is asymmetric and defined as the 16\% and 84\% percentile of the $\delta z$ posterior, we set $\sigma_z$ as the average of these values. We see that the uncertainty scales well with the noise-to-signal ratio, $1/{\rm SNR}$.}

\section{Conclusion}
\label{sec: conclusions}

In this paper, we have demonstrated the feasibility of using the \lyaf to perform clustering redshift calibration at $2<z<3$ in the context of DESI Y5 \lyaf forest data and the LSST Y10 weak lensing source sample. 
We use 10 \texttt{CoLoRe} simulations to create mocks for both tracer samples with matching noise level, number density, bias evolution, and redshift distribution over the joint DESI$\times$LSST footprint. For the \lyaf, we create different mocks with different continuum fitting methods and contamination. We demonstrate a framework to infer the biased redshift distribution for the unknown sample, $bn(z):=b(z)n(z)$, from the cross-correlation function with the binned \lyaf, $\Delta_F$, including redshift-space distortion (RSD) effects. We quantify the constraining power of this method using two models for $bn(z)$: shift mean (assuming no knowledge at $z<2$) and Gaussian Process (assuming full knowledge at $z<2$). In our baseline setting, where we use the \texttt{LyCAN} mocks, adopt a redshift bin width of $\Delta z=0.1$, and measure at the smallest angular bin at $\theta=[10.5, 13.6]$ arcmin, we achieve a $24\sigma$ detection, which yields a constraint on the mean redshift of the tomographic bin $\sigma_z/(1+\bar{z})=0.014$ for the shift mean model, and $\sigma_z/(1+\bar{z})=0.006$ for the GP model (with $\bar{z}=2$). Varying the analysis choices, we learn the following:
\begin{itemize}
    \item \lyaf continuum fitting is important for clustering redshift measurements. While the baseline using \texttt{LyCAN} works well, the conventional method, \texttt{Picca}, significantly suppresses the correlation signal by 40\% compared to the baseline. If the continuum is known exactly, the constraint can be tighter by $20\%$ compared to the baseline. 
    \item When contamination is present in the quasar spectrum, the effective \lya bias changes, and the SNR is degraded by $\sim 10\%$.
    \item Measurements at small angular scales are preferred for maximizing SNR. Adding larger scales does not further tighten the constraints on $bn(z)$.
    \item Finer redshift bin size, $\Delta z$, does not have a significant impact on the SNR or the constraints.
\end{itemize}

In this paper we only provide a proof-of-concept study.
There are caveats in the analysis presented here that can be improved in a future study. Firstly, the simulations lack power at small scales and we introduce a cut at $\theta_{\rm min}=10$ arcmin. However, smaller scales may provide higher SNR measurements and hence further improve the redshift constraint. Secondly, we used a synthetic data vector in the inference due to unexpected fluctuations with redshifts in the simulations. Both of these caveats should be addressed with improved simulations. 
Thirdly, we have not yet considered the combination of this methodology with lower-redshift clustering redshift measurements using other DESI tracers, {and we expect} that the more realistic constraints in this case should lie somewhere between the mean redshift and GP values. 
Furthermore, we do not discuss the dependence on cosmology in the model fitting. For example, the overall amplitude of the galaxy bias is completely degenerate with the $\sigma_8$ parameter. Evolution of the cosmological parameters over $2<z<3$ could also potentially bias the results.
Lastly, we have not explored bias mitigation strategies. This problem exists in conventional clustering redshift methods as well. Although several methods exist in the literature \citep[e.g.][]{2025arXiv251023565D}, further study is needed to apply them in the context of this study.

The constraining power of using \lyaf{s} to calibrate photometric galaxy redshift distributions can potentially be increased in the future. As we mentioned before, the SNR of the measurement is currently limited by the \lyaf noise. One way to reduce this noise is to combine with the measurements using other high-redshift tracers. These can be individual metal absorbers or metal forests (i.e. silicon and carbon forests) in the quasar spectra. Although these absorbers can be identified with a lower SNR or number density, they typically trace more massive structures, hence correlating better with galaxies. The other way would be to provide more \lyaf data by surveying more high-redshft quasars, or improving the SNR of the quasar spectra. This could potentially be achieved with the future spectroscopic surveys, such as the DESI extension and possibly the DESI-II programme. {Together with a large number of high-redshift Lyman-$\alpha$ emitting (LAE) galaxies  detected by DESI-II, clustering-$z$ with multiple tracers can achieve promising constraining power for Stage IV surveys.}

\lyaf also provides a way to calibrate ensemble distributions of LBGs and LAEs detected in photometric surveys. For example, LSST can detect LBGs in the redshift range $2<z<6$ via drop-outs in short wavelength bands. The method outlined in this paper can potentially constrain the redshift distribution of this population which is otherwise extremely difficult. {This could improve the robustness of high-redshift cosmological studies such as constraints on primordial non-Gaussianity \citep[e.g.][]{2025arXiv251122243P,2026MNRAS.545f2115P}.}

\section*{Acknowledgements}

The authors would like to thank Jaime Ruiz-Zapatero for his help with the Gaussian Process model.
{The authors would like to thank the anonymous referee for their constructive and insightful comments that improved the quality of this paper.}
QH and BJ acknowledge support by the ERC-selected UKRI Frontier Research Grant EP/Y03015X/1 and by STFC grant ST/W001721/1.
WT acknowledges support from the United States Department of Energy, Office of High Energy Physics under Award Number DE-SC-0011726.
W. d’A. acknowledges support from the  MICINN projects PID2019-111317GB-C32, PID2022-141079NB-C32 as well as predoctoral program AGAUR-FI ajuts (2024 FI-1 00692) Joan Oró. LC and AFR acknowledge financial support from the Spanish Ministry of Science and Innovation (MICINN) through the Spanish State Research Agency, under Severo Ochoa Centres of Excellence Programme 2025-2029 (CEX2024001442-S), the European Union (ERC Consolidator Grant, COSMO-LYA, grant agreement 101044612), and the Spanish Ministry of Science and Innovation under the PGC2021-23012NB-C41 and PID2024-159420-C41 projects. IFAE is partially funded by the CERCA program of the Generalitat de Catalunya.

\section*{Data Availability}

The codes and simulations used for this paper are available upon request.



\bibliographystyle{mnras}
\bibliography{sample631} 




\appendix

\input{conventional_cc}

\input{rsd}
\input{redshift_william}

\input{Likelihood_william}

\section{Mean subtraction and the use of randoms}
\label{sec: randoms}

\begin{figure}
    \centering
    \includegraphics[width=\columnwidth]{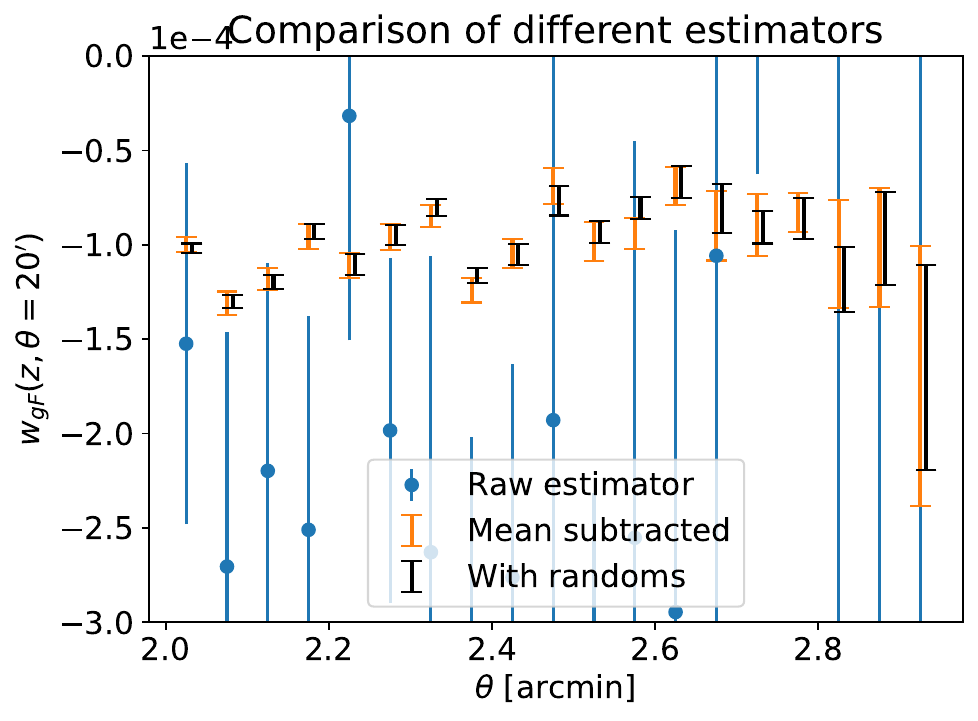}
    \caption{The measured $w_{gF}(z, \theta)$ fixed at $\theta=20$ arcmin with different estimators: the raw estimator (blue circles; Eq.~\ref{eq: w_theta}), the estimator with mean subtracted (orange lines; Eq.~\ref{eq: w_theta2}), and the estimator with randoms (black lines; Eq.~\ref{eq: rand subtraction}). The data points are slightly off-set in the $x$-axis for visual clarity. In both panels, redshift bins with $\Delta z = 0.05$ are adopted.}
    \label{fig: wsp_estimator_meansub_rand_mock_avg}
\end{figure}

We compare measurements using different estimators for $w_{\rm gF}(z, \theta)$ as mentioned in Section~\ref{sec: estimators detail}. For clarity, we only show the case with $\Delta z=0.05$, and fix the angular scale at $\theta=20$ arcmin as an example. Similar features are found for other redshift bin widths and angular scales. 

As a recap, Section~\ref{sec: estimators detail} introduced three different estimators. The raw estimator (Eq.~\ref{eq: w_theta}) computes the weighted average of the binned \lyaf flux, $\Delta_I^q$, around galaxies. The mean-subtracted estimator (Eq.~\ref{eq: w_theta2}) further subtracts the weighted mean flux over the joint footprint. Finally, the estimator with randoms (Eq.~\ref{eq: rand subtraction}) also subtracts the average flux using a uniform random catalogue over the joint footprint. 
Figure~\ref{fig: wsp_estimator_meansub_rand_mock_avg} shows the comparison of $w_{gF}(z, \theta=20')$ measured using the three estimators, averaged over 10 mock realizations. Measurements from the raw estimator show very different amplitude and scatter compared to the mean-subtracted and random estimators. The amplitude of the weighted mean flux in each redshift bin can be read off from the difference between the blue and orange data points. This is comparable to the amplitude of the cross-correlation signal, hence contributing significantly to the scatter of the raw estimator. 

The measurements from mean-subtracted and random estimators are consistent with each other, but the measurements with randoms show a slightly smaller scatter. The random catalogue is better at capturing the survey mask, and effectively cancels some noise in the estimator. We adopt the estimator with randoms as our baseline estimator.

\section{Comparison between theory and the raw measurements}
\label{sec: validation results}
\begin{figure*}
    \centering
    \includegraphics[width=\linewidth]{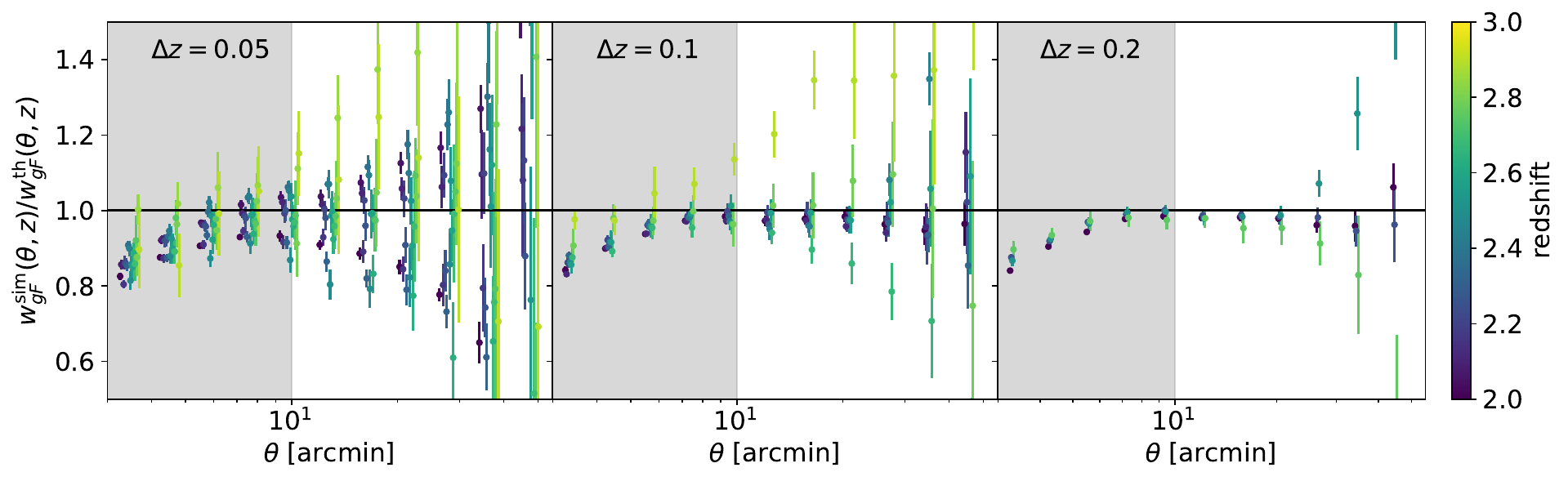}
    \caption{Comparison of the simulation measurement, averaged over 10 realizations, to the theoretical prediction of the \lya-galaxy cross-correlation function, $w_{gF}(\theta, z)$. Each panel shows measurements from different redshift bin width, $\Delta z$. Error bars are scaled for the simulation average. The gray shaded region shows the scales at which the mock deviates from linear theory. Data points at different redshifts have been slightly shifted horizontally for visual clarity.}
    \label{fig: validation_wsp_theta_z_zbins_data_over_theory_mock_avg}
\end{figure*}

Figure~\ref{fig: validation_wsp_theta_z_zbins_data_over_theory_mock_avg} shows the ratio between the measured $w_{gF}(\theta, z)$ and the linear theory prediction with $\Delta z = 0.05,0.1,0.2$. 
We see that at scales below $\theta=10$ arcmin, there is a lack of power in the simulations compared to the linear theory prediction. This is expected, because the power spectrum used in the simulation is smoothed at small scales, as mentioned in Section~\ref{sec:colore}. 

At larger scales we see reasonable agreement between the measurements and the theory. However, there seems to be a binning artifact in redshift, where, when the bin size becomes smaller, the measurements fluctuate significantly, up to $20\%$ on intermediate scales. This effect is not removed by averaging over 10 simulations. Despite running a number of tests, we could not identify the source of this fluctuation in the simulation. Hence, we decided to only use the noise level informed by these simulations, but do not directly use the data points for $n(z)$ inference. Instead, we use the synthetic data vector for all mean redshift errors quoted.


\bsp	
\label{lastpage}
\end{document}

%% file: authors.tex
\author[Hang et al.]{
Qianjun Hang,$^{1}$ \thanks{e.hang@ucl.ac.uk}
Laura Casas,$^{2}$
William d'Assignies,$^{2}$
Wynne Turner,$^{3,4}$
Andreu Font-Ribera,$^{2,5}$
\newauthor
Benjamin Joachimi$^{1}$
\\
$^1$ Department of Physics and Astronomy, University College London, Gower St, London WC1E 6BT, UK\\
$^2$ Institut de F\`{i}sica d'Altes Energies (IFAE), The Barcelona Institute of Science and Technology, Campus UAB, 08193 Bellaterra Barcelona, Spain \\
$^3$ Department of Astronomy, The Ohio State University, 4055 McPherson Laboratory, 140 W 18th Avenue, Columbus, OH 43210, USA\\
$^4$ Center for Cosmology and AstroParticle Physics, The Ohio State University, 191 West Woodruff Avenue, Columbus, OH 43210, USA\\
$^5$ Instituci\'{o} Catalana de Recerca i Estudis Avan\c{c}ats, Passeig de Llu\'{i}s Companys, 23, 08010 Barcelona, Spain\\
}

%% file: simulations_new.tex
\section{Simulations}
\label{sec:simulations}

In this section we describe the simulated data used to validate the methodology and to forecast the performance of the technique in the context of analyses of the final datasets of DESI and LSST.
We start in Section~\ref{sec:colore} by describing \texttt{CoLoRe} \citep{Ram_rez_P_rez_2022}, the software used to generate the synthetic galaxy and quasar catalogues.
In Section~\ref{sec:lyacolore} we describe \texttt{LyaCoLoRe} \citep{2020JCAP...03..068F}, the software used to generate the \lyaf fluctuations in the lines of sight towards the simulated quasars.
Finally, the \texttt{quickquasars} method used to simulate DESI spectra \citep{HerreraAlcantar2025} is summarized in Section~\ref{sec:desi}.

\subsection{Multi-tracer simulations with \texttt{CoLoRe}}
\label{sec:colore}




We generate ten $8.7\ \rm{Gpc/h}$ simulation boxes based on the {\it Planck} 2015 results \citep{Planck:2015fie} to account for cosmic variance. For the generation of the boxes we used the \texttt{CoLoRe}\xspace package described in \cite{Ram_rez_P_rez_2022}. The process begins with the generation of a Gaussian random field that simulates the matter distribution, guided by an input linear power spectrum at $z=0$, which is then evolved back in time to reach the desired redshifts. Simultaneously, the corresponding Gaussian Newtonian potential is computed.

The mock galaxies cover the redshift range $z \in [0, 3]$ and the \lyaf{s} are simulated in the range $z \in [1.8, 3]$. The mocks are generated on a low-resolution grid with cells of order $\mathcal{O}(1)\,h^{-1}\mathrm{Mpc}$. We convert the Gaussian field into a log-normal field. To mitigate extreme fluctuations that can arise from the finite grid size during the subsequent log-normal transformation, a Gaussian smoothing is applied to the power spectrum.

At this stage, QSO positions are determined by Poisson sampling of the log-normal transformed density field, using an input number density and a linear biasing model with a threshold, as described in~\cite{DESI:2025qqu}.

Finally, the radial velocity field, which accounts for the peculiar velocities required to model redshift-space distortions, is inferred from the gradient of the Newtonian potential. From the initial Gaussian density field, Gaussian density skewers are then generated from each quasar towards the center of the mock box.

\begin{figure*}
    \centering
    \includegraphics[width=0.8\linewidth]{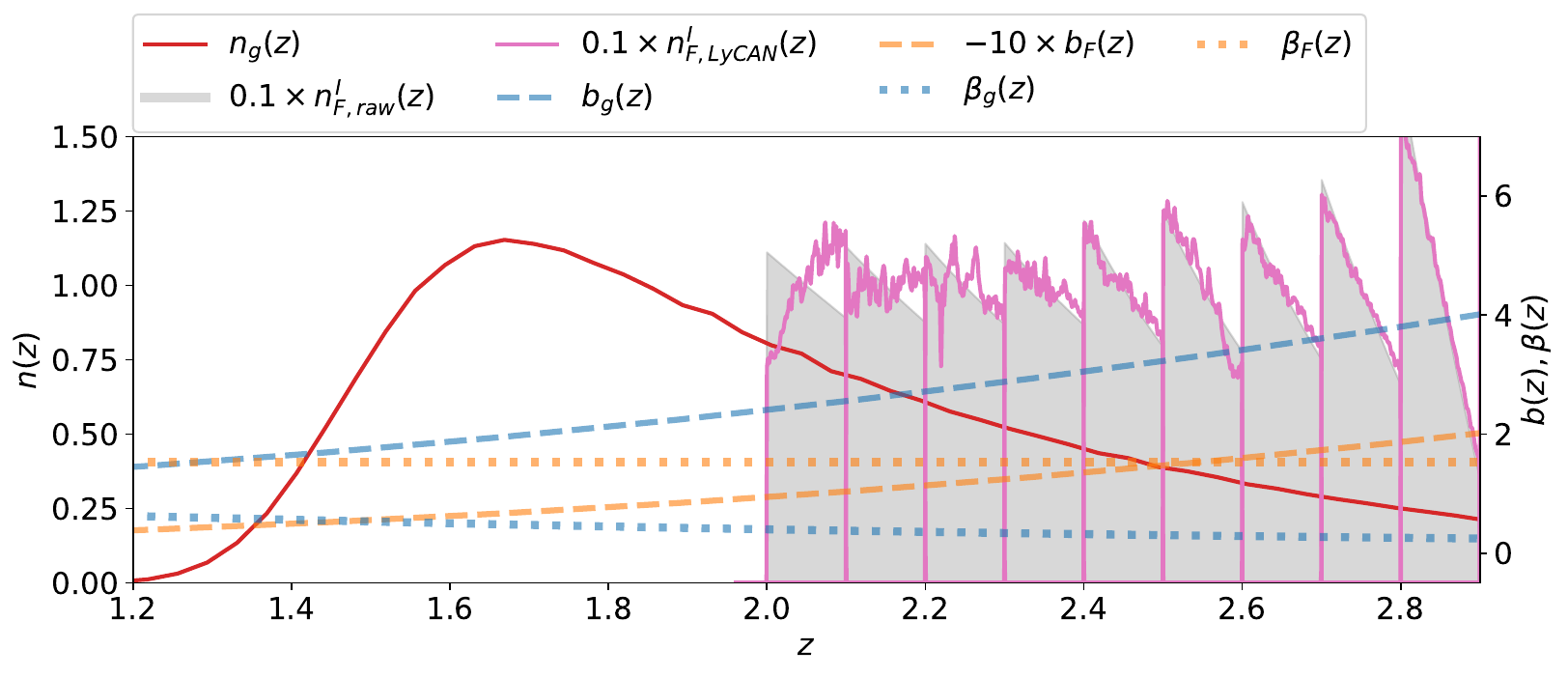}
    \caption{Redshift dependent quantities adopted in validation. The left axis shows redshift distributions of the galaxies, $n_g(z)$, and of the \lyaf using the raw mocks $n_{F, {\rm raw}}^I(z)$, and the \texttt{LyCAN} mocks $n_{F, {\rm LyCAN}}^I(z)$, with $\Delta z=0.1$ averaged over 10 mocks. The shape of $n_g(z)$ is adopted from the DESC SRD for the LSST 10-year source sample in the last tomographic bin. The $n_F^I(z)$ has been scaled by a factor of $0.1$ for visual clarity. The right ordinate shows the tracer biases, $b_{g}(z),b_F(z)$, and RSD parameters, $\beta_{g}(z),\beta_{F}(z)$, for galaxies and \lyaf, respectively. We have scaled $b_F(z)$ by a factor of $-10$ for better readability.}
    \label{fig: ng_nFI_SRD_nz_mock_avg}
\end{figure*}

We also create a mock photometric galaxy catalogue, matched to the furthest tomographic bin of the expected LSST final-year source galaxy sample. We adopt the effective number density of the `Gold' sample with $i<25.3$ to be $n^{\rm eff}_{\rm gal}=27 \, {\rm arcmin}^{-2}$ according to the LSST DESC Science Requirement Document \citep[DESC SRD][]{2018arXiv180901669T}, which incorporates the weak lensing weights described in \cite{2013MNRAS.434.2121C}. We then assume equi-populated tomographic bins, resulting in $n^{\rm eff}_{\rm gal}=5.4 \, {\rm arcmin}^{-2}$ for the highest redshift bin considered here\footnote{Note that in practice, the equi-populated tomographic bins may be assigned to the unweighted galaxy catalogue, hence the effective number density may be different in each tomographic bin.}. We also adopt the redshift distribution as described in the LSST DESC SRD.
The galaxy bias evolution takes the form of a second-order polynomial, fitted to the LSST DESC DC2 simulations \citep{2021ApJS..253...31L} with $i<25.3$. The redshift dependent quantities of the photometric galaxies are shown in Fig.~\ref{fig: ng_nFI_SRD_nz_mock_avg}. Finally, we consider a joint LSST Wide Field Survey $\times$ DESI footprint covering a total area of $5492 \,{\rm deg}^2$. The LSST mask is derived from the Rubin Operations Simulator (\texttt{OpSim}\footnote{\url{https://rubin-sim.lsst.io}}) using the simulated visits in the wide-fast-deep (WFD) region and excluding area with high galactic extinction ($E(B-V)>0.2$).

\subsection{Simulating the \lya forest with \texttt{LyaCoLoRe}}
\label{sec:lyacolore}

The next step in the mock generation process uses the \texttt{LyaCoLoRe} package, described in \cite{2020JCAP...03..068F}. This package takes the low-resolution Gaussian density skewers produced by \texttt{CoLoRe} and enhances them by adding small-scale fluctuations derived from an input one-dimensional power spectrum~\citep{McDonald_2006}. It then applies a log-normal transformation, followed by the fluctuating Gunn-Peterson approximation~\citep{Croft_1998}, which converts density fluctuations into the corresponding optical depth $\tau$. After including the redshift-space distortions computed by \texttt{CoLoRe}, the transmitted flux fraction for a quasar $q$ is obtained as $F_q(\lambda) = e^{-\tau_q(\lambda)}$.

\begin{figure*}
    \centering
    \begin{subfigure}[b]{\linewidth}
        \centering
        \includegraphics[width=\textwidth]{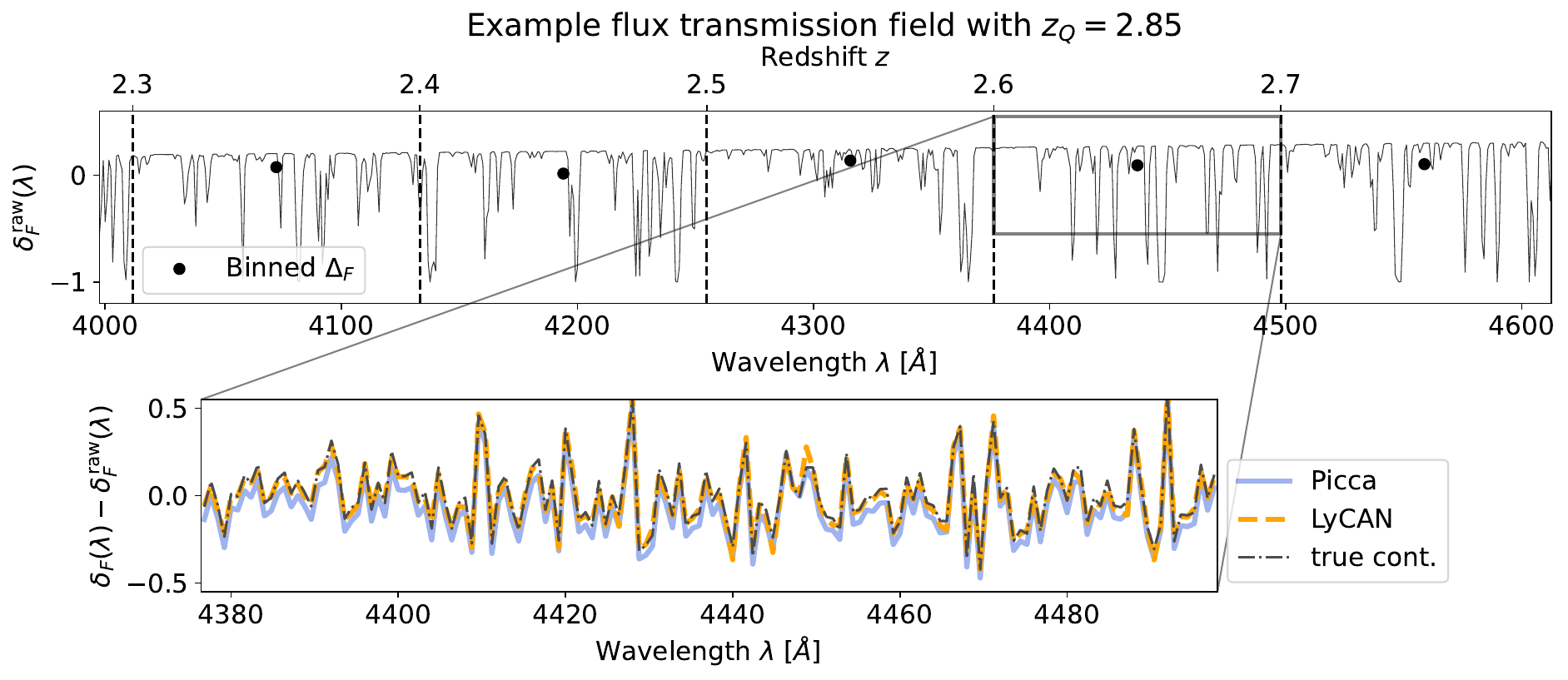}
        \caption{\lyaf flux transmission fluctuation field.}
    \end{subfigure}
    ~ 
    \begin{subfigure}[b]{\linewidth}
        \centering
        \includegraphics[width=0.7\textwidth]{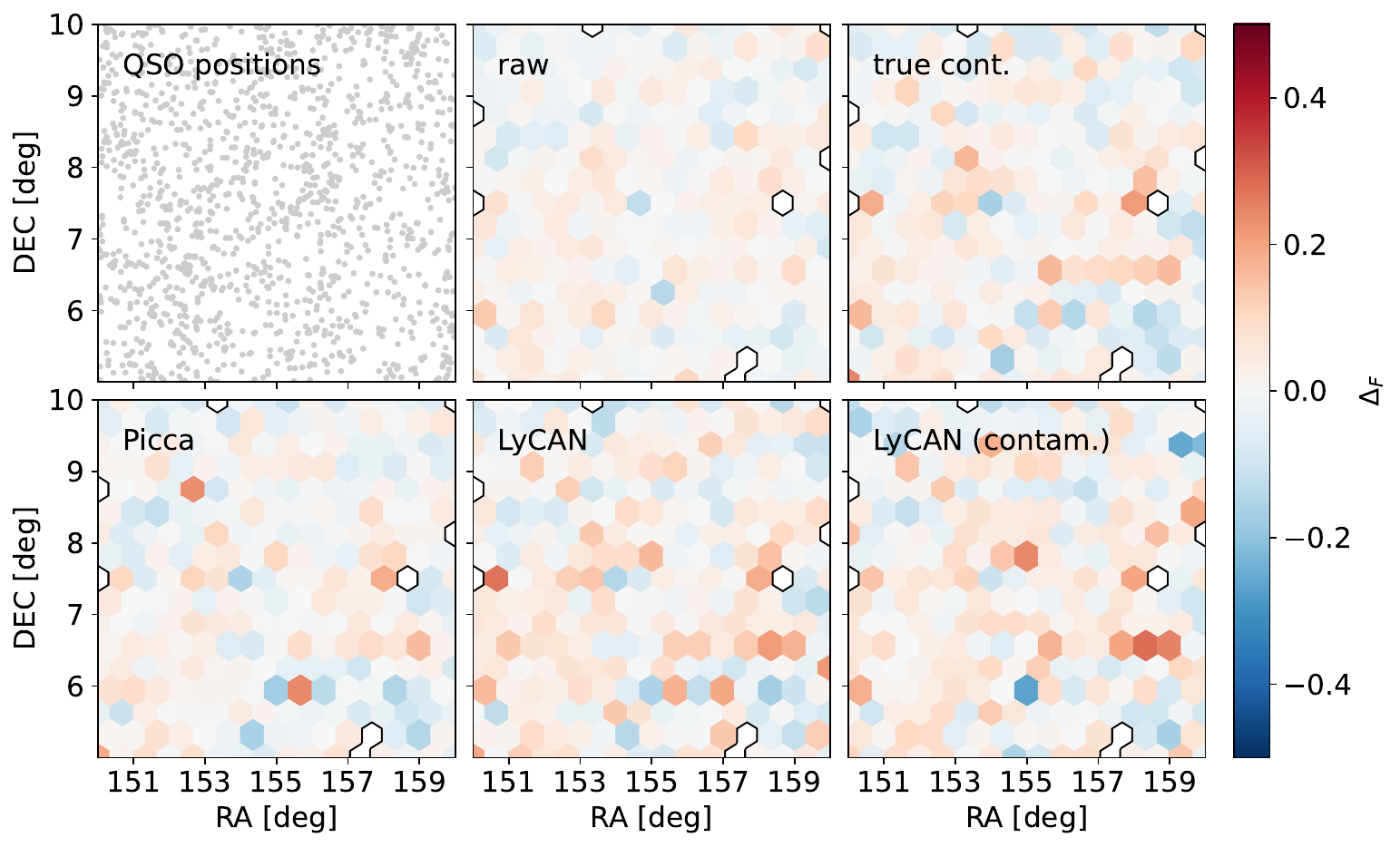}
        \caption{Binned flux with $z=[2.4,2.5]$.}
    \end{subfigure}
    \caption{\textit{Upper panel}: An example \lyaf flux transmission fluctuation field, $\delta_F^q(\lambda)$, between $3997.6 \text{\AA} <\lambda<4612.6\text{\AA}$ in the \texttt{CoLoRe} simulation for a quasar at $z_Q=2.85$. The top panel shows the `raw' flux fluctuation, which is the ground truth. The range of the wavelength shown corresponds to the minimum and maximum cuts in the quasar rest-frame, and the vertical dashed lines indicate the redshift bin edges for $\Delta z=0.1$. The black dots show the value of the binned fluctuations, $\Delta_F$ in each redshift bin. 
    The figure below shows a zoom-in of the difference between noisy $\delta_F$ and the raw analysis in one of the redshift bins. We over-plot the `true continuum' case (black dot-dashed line), the \texttt{picca} method without contamination (blue thick line), and the \texttt{LyCAN} method without contamination (orange dashed line). \textit{Lower panel}: An example of the binned flux transmission field $\Delta_F$ on a cutout of the footprint in one redshift slice $2.4<z<2.5$ for raw, true continuum, \texttt{Picca}, and \texttt{LyCAN} with and without contamination. The $\Delta_F$ values are binned spatially to reduce noise for visual clarity.}
    \label{fig: delta demo}
\end{figure*}

The resulting products at this stage are referred to as \emph{raw mocks}, since they do not yet include additional components such as quasar continua, contaminants, or instrumental effects. These elements are incorporated in subsequent processing steps. The top panel in Fig.~\ref{fig: delta demo} shows an example of $\delta_F^q(\lambda)$ for the raw analysis.

We measure and fit for the \lyaf bias in the simulations:
\begin{equation}
    b_F(z) = b_{\rm ref} \left ( \frac{1+z}{1+z_{\rm ref}} \right )^{\alpha}, 
\end{equation}
where $\alpha=2.9, b_{\rm ref} = -0.1352, z_{\rm ref} = 2.4$. We also measure the \lyaf RSD parameter $\beta_F = 1.53$, independent of redshift. These functions are shown in Fig.~\ref{fig: ng_nFI_SRD_nz_mock_avg}.

\subsection{Simulating DESI quasar spectra} 
\label{sec:desi}




After generating the quasar positions and the skewers of \lya\ transmission, we simulate DESI quasar spectra in two main steps.

First, we create DESI-like quasar populations using the \texttt{desisim} \footnote{\url{https://github.com/desihub/desisim}} package. This tool downsamples the simulated quasar catalogue to reproduce the footprint and redshift distribution of DESI Y5 quasars. Then each simulated quasar is assigned an apparent magnitude in the $r$-band and number of exposures, as described in \cite{DESI:2024txa}.

Next, we convert the \texttt{desisim} quasar catalogue into realistic DESI spectra with \texttt{quickquasars} \citep{HerreraAlcantar2025}. Noiseless spectra are generated by multiplying the post-processed transmitted flux by quasar continuum templates produced with the \texttt{SIMQSO}\footnote{\url{https://github.com/imcgreer/simqso}} package. We use the \texttt{specsim} \footnote{\url{https://github.com/desihub/specsim}} package  to add instrumental noise representative of DESI’s nominal dark-time observing conditions. These correspond to what we refer to as the \textit{uncontaminated mocks}. 

For the contaminated mocks, \texttt{quickquasars} also adds absorption features from astrophysical sources (contaminants) before applying the continuum. These include Damped \lya\ Absorbers (DLAs), Broad Absorption Lines (BALs), higher-order Lyman lines (up to L$y\epsilon$), and metal absorbers, following \cite{HerreraAlcantar2025}.

%% file: continuum_fitting_new.tex
\section{Estimating the fluctuations in the \lya forest}
\label{sec: Estimating the fluctuations}

We first briefly describe the notational convention used throughout the paper.
The DESI spectra are discretized into pixels of width $0.8 \text{\AA}$. 
Hereafter we shall refer to a specific pixel by subscript $i$. 
We refer to any function of wavelength $f(\lambda)$ to be implicitly discretized in the simulation, i.e. $f(\lambda_i)$ or $f_i$. 
We use the index $q$ to refer to a quantity or function associated with a quasar $q$. 
We shall refer to a `line-of-sight' or a particular direction $\hat{\mathbf{n}}$ interchangeably as associated with a quasar $q$.
Later on we shall introduce the binned flux transmission fluctuation, $\Delta_{I}^q$, where we refer to the particular bin number as subscript $I$.

Given the observed flux at a given pixel $f_q(\lambda_i)$, we would like to estimate the fluctuation in the \lya transmitted flux fraction:
\begin{equation}
    \hat \delta_q(\lambda_i) 
        = \frac{f^q_i}{\hat C^q_i \hat{\bar{F}}_i} - 1 ~,\label{eq: transmission_field}
\end{equation}
where $\hat C^q_i$ is an estimate of the (unknown) quasar continuum and $\hat{\bar{F}}$ an estimate of the mean transmitted flux fraction.

The process to estimate both $\hat C^q_i$ and $\hat{\bar{F}}_i$ is known as \textit{continuum fitting}, and there are several methods proposed in the literature.
In this publication we use two different methods: on the one hand, we use \texttt{Picca}, the continuum fitting code used in \lya BAO analysis of the eBOSS and DESI collaborations \citep{dMdB2020, DESI_DR1_Lya_BAO, DESI_DR2_Lya_BAO}.
On the other hand we use \texttt{LyCAN} \citep{2024ApJ...976..143T}, a machine learning-based method that uses the observed flux outside of the \lya forest region to predict the unabsorbed quasar continuum in the \lya region. 

In order to show the maximum performance that we could hope to achieve with the clustering redshift technique, in the following sections we will also discuss the results when using the \textit{true continuum} used in the simulations, avoiding this way the noise added in continuum fitting.

\subsection{Picca}

In \lyaf\ analyses, we need to determine the non-fluctuating component of the forest, $\hat{C}^q_i \hat{\bar{F}}_i$ in Eq.~\ref{eq: transmission_field}, in order to extract the fluctuating field $\hat{\delta}_q(\lambda_i)$ of interest. However, this quantity cannot be known a priori, so we rely on an estimate, which  introduces distortions into the $\delta$ field.

For the estimate of the continuum, \texttt{Picca} assumes a universal template that depends on rest-frame wavelength, $T(\lambda_{\mathrm{rest}})$. Each forest is then fit with the form
\begin{equation}
    (a_q +b_q\Lambda_{q\lambda})T_{\lambda z_q}
\end{equation}
where $\Lambda = \log(\lambda) - \overline{\log(\lambda)}$, and $T_{\lambda z}$ represents the template shifted to the redshift of the quasar.

However, as a consequence of this procedure, we obtain
\begin{equation}
    \hat \delta_q^m(\lambda_i) = \frac{1 + \hat \delta^t_q(\lambda_i) + n}{1 + \epsilon_q} -1
\end{equation}
where $\epsilon_q$ represents the distortion introduced by the estimation, $n$ is a noise factor and the labels $m$ and $t$ are to indicate the measured and true transmission fields.
For further details, see \cite{Busca:2025knm}; the key idea is that $\epsilon_q$ depends on the true transmission field across all wavelengths, thereby coupling small and large scales.





When we apply this procedure to the mocks we refer to them as \textit{Picca mocks}, in comparison to the other continuum fitting method. When instead we remove the real continuum which is known in the mocks case we refer to these as \textit{true continuum mocks}. The zoom-in of the top panel in Fig.~\ref{fig: delta demo} shows examples of how the true continuum (black dot-dashed line) and \texttt{Picca} (blue line) mocks differ from the raw analysis. While \texttt{Picca} recovers well the high-frequency fluctuations, over large wavelength range, it produces a systematic offset from the true continuum case.

\subsection{LyCAN}

The \lya Continuum Analysis Network (\texttt{LyCAN}) is a convolutional neural network designed to predict the unabsorbed continuum independently of the forest region \citep{2024ApJ...976..143T}. It predicts the continuum in the rest-frame wavelength range $1040-1600$ \AA~based only on the red side of the \lya emission line ($1216-1600$ \AA). The training data consist of a set of synthetic spectra derived from high-resolution, low-redshift Hubble Space Telescope/Cosmic Origins Spectrograph (COS) observations, supplemented by an equal number of DESI Y5 mock spectra. The method achieves a median error of $1.5\%$ on the DESI mock spectra, and $4.1\%$ on the original COS spectra.

When applied to the DESI Year 1 spectra to measure the evolution of the effective optical depth, \texttt{LyCAN} yields results in good agreement with high-resolution ground-based measurements around $z = 2$, demonstrating the reliability of its continuum predictions. Moreover, because \texttt{LyCAN} predicts the continuum independently of the forest region, it preserves large-scale information that is lost in traditional \texttt{Picca} continuum fitting and avoids distortions to the measured correlation function \citep{2025arXiv250914322T}.

We refer to this set of mocks as the \textit{LyCAN mocks}. Additionally, we make \textit{Contaminated LyCAN mocks}, which also include  contamination as described in Section.~\ref{sec:desi} before applying the \texttt{LyCAN} continuum fitting. Fig.~\ref{fig: delta demo} shows an example of how the \texttt{LyCAN} mock (orange dashed line) differs from the raw analysis in the zoom-in of the top panel. \texttt{LyCAN} can recover the true continuum mock (black dot-dashed line) well on both large and small wavelength intervals.

\subsection{Pixel weights}

Following the \lya BAO analyses of DESI \citep{DESI_DR1_Lya_BAO, DESI_DR2_Lya_BAO}, we define per-pixel weights that will be used to measure the correlations:
\begin{equation} \label{eqn:weights}
 W_q(\lambda) = \frac{1}{\sigma^2_{{\rm tot}, q}(\lambda)} ~,
\end{equation}
where $\sigma^2_{\rm tot}$ is the total pixel variance, defined as
\begin{equation}
\sigma^2_{{\rm tot}, q}(\lambda) = \eta_{\rm pip}(\lambda) \left( \frac{\sigma_{{\rm pip}, q}(\lambda)}{\overline{F}C_{q}(\lambda)}\right)^2 +  \eta_{\rm LSS} \, \sigma^2_{\rm LSS}(\lambda) ~,
\end{equation}
with $\sigma^2_{\rm pip}$ is the instrumental noise variance, $\sigma^2_{\rm LSS}$ the intrinsic variance of the \lya forest fluctuations, and the functions $\eta_{\rm pip}(\lambda)$ and $\eta_{\rm LSS}(\lambda)$ modify the contributions from each term to the weights \citep[see][for a detailed description of the \lya weights]{DESI_DR1_Lya_BAO}.
In \lya BAO analyses, where the correlations are averaged over a wide redshift range ($2 < z < 4$), it is common to further modify the weights with a multiplicative correction to up-weight pixels at higher redshift (longer wavelenghts) where the \lya clustering is stronger. 
{However, since here we will measure correlations using very narrow redshift bins, the evolution of the cross-correlation signal within the bin is negligible. Hence, we do not apply this additional redshift-dependent modification to the weights as done in the BAO analysis.}
For the raw analysis where no noise is present, all pixels receive the same weight. 

In order to make sure that the fluctuations have zero mean when averaged with the same weights that will be used in the measurement of the correlations, we apply a final correction and divide $\left(1+\hat \delta_q(\lambda) \right)$ by its weighted mean, defined as:
\begin{equation}
 \left< 1 + \hat \delta_i \right> = 1 + \frac{1}{\sum_q W^q_i} \sum_q W^q_i ~ \hat \delta^q_i ~,
 \label{eq: 1+delta weight}
\end{equation}
with $W^q_i=W_q(\lambda_i)$.
When computing this mean, and when computing the correlations in the next sections, we limit our analysis to pixels in the rest-frame wavelength range $[\lambda^r_{\rm min}, \lambda^r_{\rm max}] = [1040\text{\AA}, 1200\text{\AA}]$.


\subsection{Rebinning the \lya fluctuations}

The native (line-of-sight) pixelization of \lya data has very high resolution when compared to the typical width of the tomographic bins in photometric galaxy surveys.
For instance, the size of a pixel in a DESI spectrum is 0.8 \AA, corresponding to a redshift resolution of $\Delta z = 0.000658$, independent of redshift.

To construct the reference sample for clustering redshifts, we split the redshift (wavelength) of the transmission field into $N$ bins in range $2<z<3$. The wavelength-redshift relation is simply given by $z = (\lambda - \lambda_0)/\lambda_0$.
For each redshift bin $I$, we identify all pixels within the bin, $i\in I$ for a given quasar, and compute the re-binned flux transmission fluctuation for each line of sight:
\begin{equation}
    \Delta^q_I = \frac{\sum_{i \in I} W_i^q \hat \delta_i^q}{\sum_{i \in I} W^q_i} ~, 
    \label{eq: binned delta f}
\end{equation}
where the weights $W^q_i$ are defined above for the respective cases. 
Furthermore, we assign each binned transmission fluctuation $\Delta^q_I$ a weight given by the sum of the pixel weights in this bin, i.e.,
\begin{equation}
    W^q_I = \sum_{i \in I} W^q_i ~.
\end{equation}

We can refer to the ensemble of these binned flux transmission fluctuations as the \textit{catalogue} of the flux transmission field, $\Delta_F(\hat{\mathbf{n}}, z_I)$, where $\hat{\mathbf{n}}$ is the angular position of the source quasar on the sky. The lower panel of Fig.~\ref{fig: delta demo} shows examples of this field with different forest configurations for one of the \texttt{CoLeRe} simulations. The top left panel shows the discrete backlight QSO positions. For better visual effects, we binned $\Delta_F(\hat{\mathbf{n}}, z_I)$ spatially into hexagon cells for all other panels. Empty cells receive no \lyaf. The panels show how $\Delta_F(\hat{\mathbf{n}}, z_I)$ differ for the raw, true continuum, \texttt{Picca}, \texttt{LyCAN}, and contaminated \texttt{LyCAN} mocks respectively. 

The equivalent expression for the galaxy catalogue is the projected galaxy density contrast field:
\begin{equation}
    \Delta_{g}(\hat{\mathbf{n}}) = \frac{N_g(\hat{\mathbf{n}}) - \bar{N}_g}{\bar{N}_g},
    \label{eq: Delta g}
\end{equation}
where $N_g(\hat{\mathbf{n}})$ is the observed galaxy number at the angular direction $\hat{\mathbf{n}}$ and $\bar{N}_g$ is the mean observed galaxy number over the galaxy survey footprint. Notice that in practice, there can be weights in this quantity, either modifying the observed galaxy number on the RHS, or directly applied to the LHS of Eq.~\ref{eq: Delta g}. {In our simulation setup we use uniform weights for the galaxy catalogue.}

In practice, when we measure the two-point angular correlation function, we do not use Eq.~\ref{eq: Delta g} directly. Rather, we use estimators detailed in Sec.~\ref{sec: estimators}.

%% file: conventional_cc.tex
\section{Conventional clustering redshift methods}
\label{sec: old cc}

We briefly review the conventional clustering redshift method and its assumptions below. We shall subscripts `r' and `u' to denote reference and unknown samples. The projected tracer density contrast can be written as:
\begin{equation}
    \Delta_{\rm r,u}(\hat{\mathbf{n}}) = \int_0^{\infty}\dz\, b_{\rm r,u}(z)\, n_{\rm r,u}(z) \,\delta^{\rm 3D}_{\rm m}(\mathbf{x}, z),
\end{equation}
where the 3D position vector is given by $\mathbf{x} = (\chi(z), \chi(z)\hat{\mathbf{n}})$, $b_{\rm r,u}(z)$ is the linear reference/unknown sample galaxy bias, and $n_{\rm r,u}(z)$ is the redshift distribution, normalized by $\int \dz \,n_{\rm r,u}(z)=1$. $\delta^{\rm 3D}_{\rm m}(\mathbf{x}, z)$ is the 3-dimensional matter density fluctuation at comoving distance $\chi(z)$ and angular position $\hat{\mathbf{n}}$, at redshift $z$. The angular cross-correlation function can be written as:
\begin{align}
      w_{{\rm ru}}(\theta) = \int_0^{\infty} \dz_1 \int_0^{\infty} \dz_2 \, b_{\rm r}(z_1)\, & b_{\rm u}(z_2)\, f(z_1, z_2)\, \nonumber \\
      & \times \xi_{\rm m}[r(\theta, z_1, z_2), z_1, z_2],
    \label{eq: w}
\end{align}
where $f(z_1, z_2)$ is the joint distribution of reference - unknown pairs, and $\xi(r, z)$ is the 3D matter correlation function depending on the pair redshift and the separation of the pair
\begin{equation}
    r(\theta, z_1, z_2) = \sqrt{[\chi(z_2) - \chi(z_1)]^2 + [f_K(z_1)\theta]^2}.
    \label{eq: R}
\end{equation}
In the above equation, $f_K(z)$ is the comoving angular diameter distance, equal to $\chi(z)$ for a flat universe. 
The joint distribution in redshifts can be written as $f(z_1, z_2) = n_{\rm r}(z_1) n_{\rm u}(z_2)$.
Up to this point, the 3D correlation function contains all effects and this is the same as in Section~\ref{sec: theory}.

The conventional clustering redshift method makes two main simplifying assumptions. The first assumption is that redshift-space distortion (RSD) is negligible in the reference sample, such that one can use the real-space correlation function in Eq.~\ref{eq: w}. Given that linear RSD mainly affects the bin edge, this assumption is valid if the redshift bins are wide enough\footnote{Alternatively, one could work with smaller angular scales, but the `fingers-of-god' non-linear RSD effect becomes important below certain angular scales.} such that the edge effects can be ignored. 
The second assumption is that reference-unknown pairs across the reference bin edges can be ignored, and only pairs from the overlapping redshift range contribute to the correlation function. This is the Limber approximation, equivalent to adding a Dirac delta function to the joint distribution, i.e., $f(z_1, z_2 ) = n_{\rm r}(z_1) n_{\rm u}(z_2) \delta^D(z_1- z_2)$ \citep[c.f. Eq.~10 of][]{Leonard_2023}. 
This assumption is valid again if the bin is wide and (in general) the angular scale is small (see Appendix~\ref{sec: Modeling and Limber approximations}). These assumptions are in general true for low redshift clustering redshift measurements \citep[e.g.][]{van_den_Busch_2020,2022MNRAS.510.1223G}.
This gives:
\begin{equation}
    w_{\rm ru}(\theta) = \int_0^{\infty} \dz \, b_{\rm r}(z)\, b_{\rm u}(z)  \,n_{\rm r}(z) \,n_{\rm u}(z)\, \xi_{\rm m}(R[\theta, z], z).
    \label{eq: w2}
\end{equation}


To further simplify the integral, one can assume that the reference sample has a thin top-hat redshift distribution with width $\Delta z$, such that one can replace the integral limit in Eq.~\ref{eq: w2} with $[z_I-\Delta z/2,z_I+\Delta z/2]$. We also assume that the bias and the redshift distribution of the tracer samples do not change significantly over $\Delta z$, such that:
\begin{align}
    w_{\rm ru}(\theta, z_I) & \approx b_{\rm r}(z_I)\, b_{\rm u}(z_I)\, n_{\rm u}(z_I) \int_{z_I-\Delta z/2}^{z_I+\Delta z/2} \dz_2 \, \xi_{\rm m}(r, z_I) \nonumber\\
    &= b_{\rm r}(z_I)\, b_{\rm u}(z_I)\, n_{\rm u}(z_I)\,w_{\rm m} (\theta, z_I).
    \label{eq: w3}
\end{align}
Here we explicitly included $z_I$ in the argument of $w_{\rm ru}$, and we rewrite the integral as the dark matter 2D angular correlation function at redshift $z_I$, $w_{\rm m}(\theta, z_I)$.

A similar expression can be derived for the auto-correlation of the reference/unknown sample in the same top-hat bins:
\begin{equation}
    w_{\rm rr,uu}(\theta, z_I) = \frac{1}{\Delta z}b_{\rm r,u}^2(z_I) w_{\rm m}(\theta, z_I).
\end{equation}
Hence, one can construct an estimator for the unknown redshift distribution via:
\begin{equation}
    n_{\rm u}(z_I) = \frac{w_{\rm ru} (\theta, z_I)}{\Delta z\sqrt{  \,w_{\rm rr} (\theta, z_I) w_{\rm uu}(\theta, z_I)}}.
    \label{eq: nu 2}
\end{equation}
In practice, it is difficult to measure $w_{\rm uu}(\theta, z_I)$ in thin top-hat bins, hence there is typically some model dependence in the $w_{\rm uu}(\theta, z_I)$ term (either in terms of bias evolution, or in terms of the spread of photo-$z$). For more details, see \cite{2025A&A...702A.155D}.

\begin{figure}
    \centering
    \includegraphics[width=\linewidth]{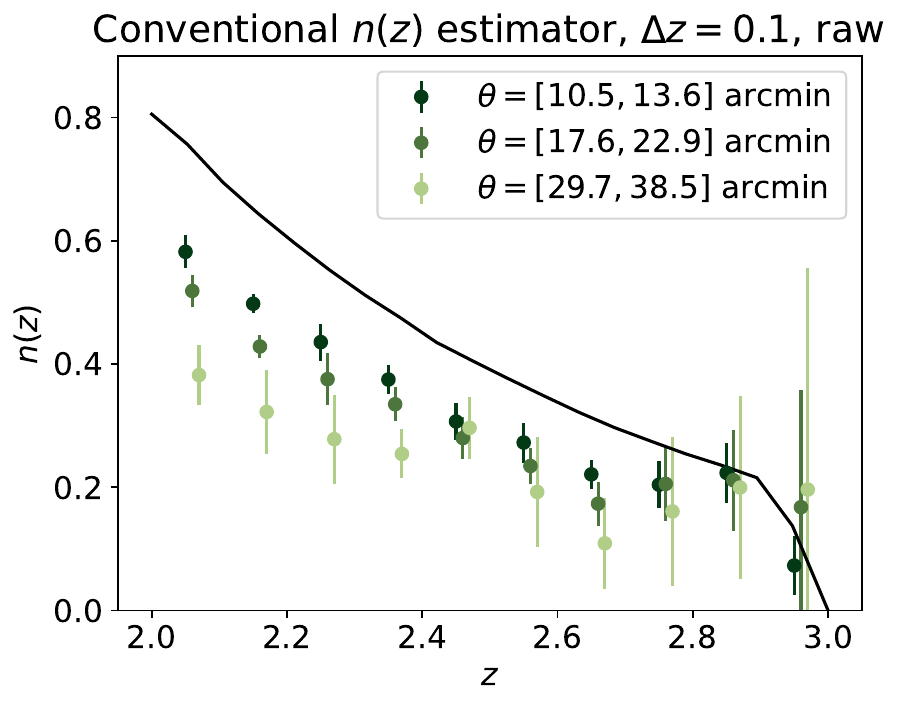}
    \caption{The conventional $n(z)$ estimator applied to the \lyaf using the raw mocks and $\Delta z=0.1$. The black line shows the true unknown $n(z)$, and the data points show the simulation measurements, averaged over 10 realizations, and measured at three angular scales. The deviations mainly comes from RSD effects in the \lya and galaxy auto-correlations.}
    \label{fig: conventional_nz_methods}
\end{figure}

Some assumptions here do not hold at high redshifts. Naively applying the estimator in Eq.~\ref{eq: nu 2} results in a biased estimate of the redshift distribution as shown in Fig.~\ref{fig: conventional_nz_methods}. We provide some explanations in Appendix~\ref{sec: Modeling and Limber approximations}.


%% file: rsd.tex
\section{Linear redshift-space distortion}
\label{sec: rsd}

In this section we briefly review the linear redshift-space distortion (RSD) effect, and provide expressions for the 3D cross-correlation function in redshift space. The derivation comes from the point of view of galaxy clustering, but \lyaf has a very similar expression.
The projected density fluctuation for a tracer $t$ in redshift space, $\Delta_t^s(\hat{\mathbf{n}})$, is given by 
\begin{equation}
    1 + \Delta_t^s( \hat{\mathbf{n}}) = \int \mathrm{d}\chi \, f(s) [1+ \delta_t^{3D}(\mathbf{x}, z)],
\end{equation}
where $\mathbf{x} = (\chi, \chi\hat{\mathbf{n}})$ is the 3D coordinate, and $s$ is the redshift space comoving distance: $s = \chi + \tilde{\mathbf{v}}\cdot\hat{\mathbf{n}}$, with $\tilde{\mathbf{v}} = \mathbf{v}/aH$, and $\mathbf{v}$ is the velocity vector of the tracer, $a$ is the expansion factor and $H$ is the Hubble parameter. $f(s)$ is the weighting function along the line of sight (here related to the redshift distribution of the tracer via $f(s)ds = n(z)dz$).
One can expand $f(s) = f(\chi) + (\mathrm{d}f/\mathrm{d}\chi)\tilde{\mathbf{v}}\cdot\hat{\mathbf{n}}$ and write $\Delta_t^s( \hat{\mathbf{n}}) = \Delta_t^r( \hat{\mathbf{n}}) + \Delta_t^{\rm RSD}(\hat{\mathbf{n}})$. The RSD correction term is given by:
\begin{equation}
    \Delta_t^{\rm RSD}(\hat{\mathbf{n}}) = \int \mathrm{d}\chi \, \frac{\mathrm{d}f(\chi)}{\mathrm{d}\chi} \tilde{\mathbf{v}}\cdot\hat{\mathbf{n}} = - \int \mathrm{d}\chi \, f(\chi) \frac{\partial [\tilde{\mathbf{v}}(\chi, \chi\hat{\mathbf{n}},z) \cdot\hat{\mathbf{n}}]}{\partial\chi},
    \label{eq: delta rsd}
\end{equation}
where the second equality is given by integration by parts. 
The peculiar velocity is related to density fluctuation at linear order in Fourier space \citep{1987MNRAS.227....1K}:
\begin{equation}
    \tilde{v}_{\parallel}(\mathbf{k},z)\equiv \tilde{\mathbf{v}}(\mathbf{k},z) \cdot \hat{\mathbf{n}} = i \frac{\beta_t(z)}{a(z)H(z)}  \frac{\mu}{k} \delta^{3D}_{t}(\mathbf{k},z),
\end{equation}
where $\mathbf{k}$ is the 3D wave vector, $|\mathbf{k}|=k$, and $\mu = \mathbf{k}\cdot\hat{\mathbf{n}}/k$ is the cosine angle between the velocity and the line-of-sight. For galaxies, $\beta_t = f/b_t$ is the distortion parameter, and $f=\mathrm{d}\ln D/\mathrm{d}\ln a$ is the linear growth rate. For \lyaf, $\beta_t$ is left as a free parameter.
The derivative along line of sight in Eq.~\ref{eq: delta rsd} in Fourier space brings down a factor of $ik\mu$, making
\begin{equation}
    \Delta_t^s(\hat{\mathbf{n}}) = \int \mathrm{d}z \, n_t(z) b_t(z) \frac{1}{\sqrt{2\pi}} \int \mathrm{d}^3\mathbf{k}\, e^{i\mathbf{k}\cdot\mathbf{x}}(1+ \beta_t (z) \mu^2 )\delta^{3D}_{\rm m}(\mathbf{k},z).
\end{equation}
The above expression is the standard RSD formalism. The inclusion of $\mu$ makes the expression in the integral anisotropic. 


The cross-correlation of the two tracers $({\rm r}, {\rm u})$ in redshift space is given by
\begin{equation}
    \langle \Delta^{s}_{\rm r} (\nvec)\Delta_{\rm u}^s (\nvec)\rangle =  \int \mathrm{d}z \int \mathrm{d}z'\,  n_{\rm r}(z) n_{\rm u}(z') \xi_{\rm ru}(\theta, z,z'),
\end{equation}
where
\begin{align}
    \xi_{\rm ru}(\theta, z,z') = b_{\rm r}(z) b_{\rm u}(z') \int \mathrm{d}^3\mathbf{k}\, e^{i\mathbf{k}\cdot{\mathbf{x}}}&[1+\beta_{\rm r}(z)\mu^2] \nonumber\\
    & \times [1+\beta_{\rm u}(z')\mu^2]P(k, \bar{z}).
\end{align}
$P(k,\bar{z})$ is the 3D matter power spectrum at the mean redshift of the pair $\bar{z}$, and it is obtained from $\langle \delta_{\rm m}(\mathbf{k}) \delta_{\rm m}(\mathbf{k}')\rangle = P(k)\delta^D(\mathbf{k} - \mathbf{k}')$.
The above integral can be written in terms of the 3D real-space correlation function $\xi(r,\bar{z})$ as the Fourier transform of $P(k,\bar{z})$:
\begin{align}
    \xi_{\rm ru}(\theta, z,z') = b_{\rm r}(z) b_{\rm u}(z') [\xi_0(r,\bar{z}) P_0(\mu) +  \xi_2(r,\bar{z}) P_2(\mu)   \nonumber\\
    + \xi_4(r,\bar{z}) P_4(\mu)],
    \label{eq: multipole}
\end{align}
where $P_{\ell}(\mu)$ is the Legendre polynomial of order $\ell$, and $\xi_{\ell}(r)$ are the multiple moments of $\xi(R)$. 
Below we omit the dependence on $z$ for simplicity:
\begin{align}
    \xi_0(r) & = \left(1+\frac{1}{3}(\beta_{\rm r} + \beta_{\rm u}) + \frac{1}{5}\beta_{\rm r}\beta_{\rm u}\right) \xi(r),\\
    \xi_2(r) & = \left( \frac{2}{3}(\beta_{\rm r} + \beta_{\rm u}) + \frac{4}{7} \beta_{\rm r}\beta_{\rm u} \right)[\xi(r) - \bar{\xi(r)}],\\
    \xi_4(r) & = \frac{8}{35} \beta_{\rm r}\beta_{\rm u} \left[\xi(r) + \frac{5}{2}\bar{\xi}(r) - \frac{7}{2} \bar{\bar{\xi}}(r) \right],
\end{align}
and
\begin{align}
    \bar{\xi}(r) & \equiv \frac{3}{r^3} \int_0^r \mathrm{d}r'\, r'^2\, \xi(r'),\\
    \bar{\bar{\xi}}(r) & \equiv \frac{5}{r^5} \int_0^r \mathrm{d}r'\, r'^4\, \xi(r').
\end{align}
Here, $r(\theta, z, z')$ is defined in Eq.~\ref{eq: R}, and $\mu=[\chi(z) - \chi(z')]/r$ is the cosine angle between the pair separation and the line of sight.

%% file: redshift_william.tex
\section{The validity of conventional clustering redshift approximations at high redshift}
\label{sec: Modeling and Limber approximations}

\subsection{The top-hat approximation for the reference sample}\label{app:tophat}

\begin{figure*}
    \centering
    \includegraphics[width=\linewidth]{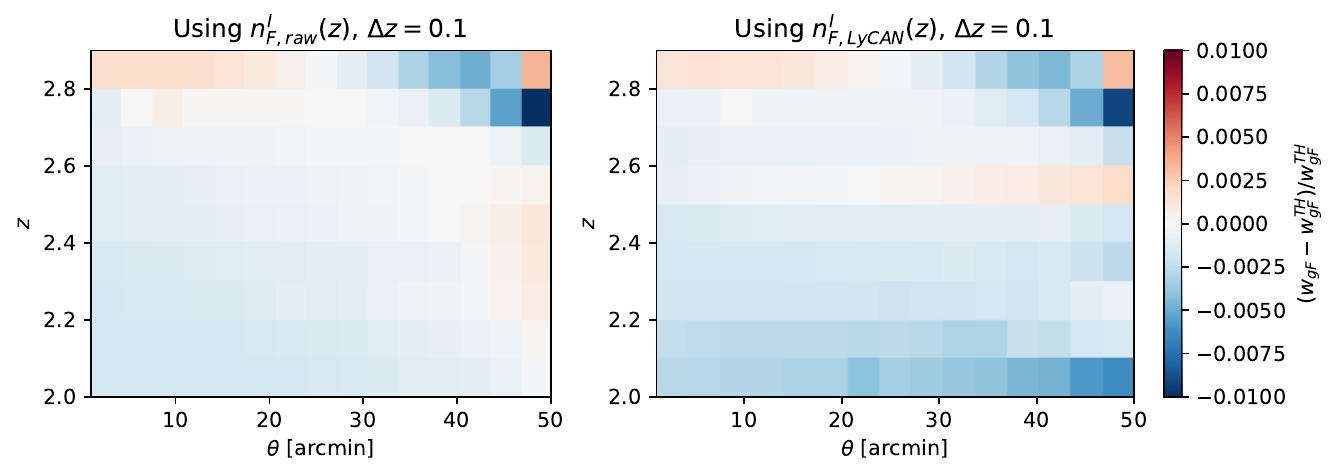}
    \caption{Fractional difference in the \lya-galaxy angular cross-correlation using the binned \lyaf redshift distribution $n_F^I(z)$ and a top-hat distribution, as a function of angular scale and redshift with redshift bin size $\Delta z=0.1$. The left panel uses the $n_F^I(z)$ from the raw analysis, and the right panel uses that from the \texttt{LyCAN} reconstruction. In general the difference is less than $1\%$.}
    \label{fig:tophat_effect}
\end{figure*}

Given that the spectroscopic reference sample is split into thin redshift bins, their redshift distributions are often assumed to be top-hat functions for computational efficiency. Here we quantify the error made in this assumption compared to using the actual redshift distributions. Here we consider the raw analysis distribution, $n_{{ F}, {\rm raw}}^I(z)$, and the \texttt{LyCAN}-reconstructed distribution, $n_{{ F}, {\rm LyCAN}}^I(z)$, both shown in Fig.~\ref{fig: ng_nFI_SRD_nz_mock_avg}. 

Figure~\ref{fig:tophat_effect} shows the fractional difference in the theoretical \lya-galaxy angular cross-correlation between using the full $n_F^I(z)$ and using a top-hat function with $\Delta z=0.1$. In most cases, the difference is less than $1\%$, with a larger deviation at the highest redshift bin, where the shape of $n_F^I(z)$ deviates noticeably from a top-hat.

For completeness, we also check the same case for auto-correlations. We observe a very similar trend where most deviation is sub-$1\%$, although the sensitivity to the actual shape is enhanced due to the quadratic dependence of the $n(z)$ shape. 
We conclude that we can safely use the top-hat redshift distribution for the reference sample here. 




\subsection{Limber approximation}

In this section we discuss the validity of the Limber approximation, i.e. the difference between using Eq.~\ref{eq: w} and Eq.~\ref{eq: w2}. We do not consider RSD effect. 
For clustering redshifts, the Limber approximation refers to the approximation 
\begin{equation}
    \int_{z_i\pm\Delta z/2} \diff z_1 \int  \diff z_2\, f(z_1,\,z_2,\,\theta) \approx  \Delta z\int  \diff z\, f(z_i,\,z,\,\theta), \label{eq:limber_approx}
\end{equation}
where the second integral domain  is $z_i\pm \Delta z/2$ for auto-correlations, and $0<z<\infty$ for cross-correlation, and where $f$ involves pair counts. Typically, in this work $f(z_1,z_2,\theta)=n_x(z_1)\,n_y(z_1)\,b_x(z_1)\,b_y(z_2)\,\xi_{\rm m}(z_1,z_2,\theta)$. Notice that  assuming a top-hat distribution for the reference sample, \textit{cf.} App. \ref{app:tophat}, we get rid of the $\Delta z $ of the right hand side. 
Taking the right term, it is usually  further assumed
\begin{equation}
    \int  \diff z\, n_y(z)\,b_y(z)\,\xi_{\rm m}(z_i,z,\theta) \approx  \,n_y(z_i)\,b_y(z_i)\int  \diff z\,\xi_{\rm m}(z_i,z,\theta).\label{eq:limber_approx2}
\end{equation}
 The usual justification being that  $\xi_{\rm m} (z_i,z,\theta)$ is non-zero only for $z-z_i$ `close to 0', so that redshift evolution can be neglected. 
 
\begin{figure*}
    \centering
    \includegraphics[width=1\linewidth]{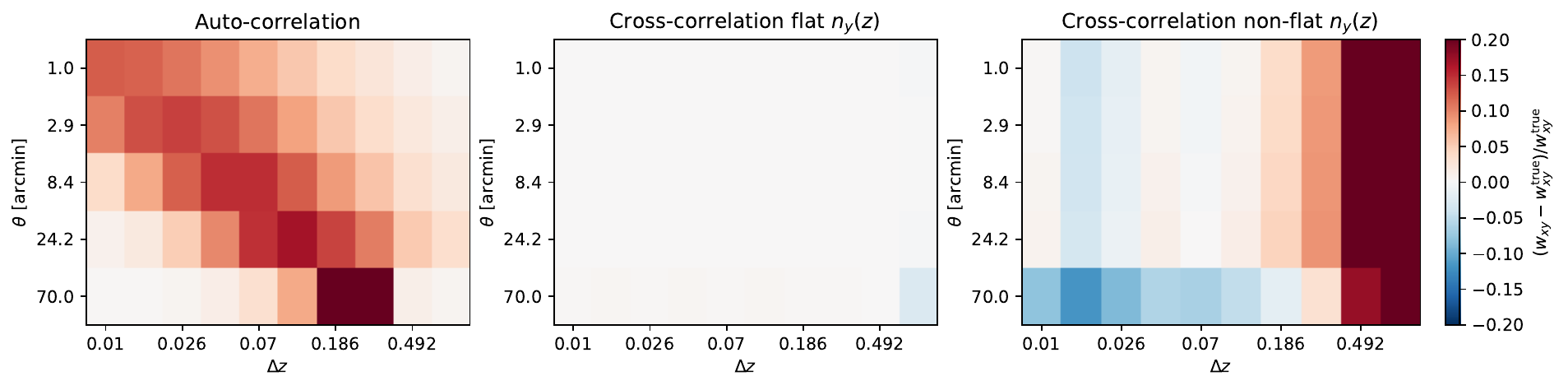}
    \caption{Deviations of the approximation \eqref{eq:limber_approx} or \eqref{eq:limber_approx2} versus the exact correlation, for the auto-correlation (left), cross-correlation with flat $n_y(z)$ (middle), and cross-correlation with a non-flat $n_y(z)$ (right, Gaussian centered at $z=2.45$, with $\sigma=0.2$). We plot the deviation as function of the angular scale, and redshift width of the top-hat bin $n_x(z)$, centered at $z=2.5$.}
    \label{fig:limber_approx}
\end{figure*}

 We start by evaluating the approximation of Eq. \eqref{eq:limber_approx}, for auto-correlations, by taking a top-hat bin for $n_{x,y}$ centered at $z=2.5$, with width $\Delta z$, and a galaxy bias fixed to unity. Then we evaluate the approximation for cross-correlations, with a top-hat $n_x$, a flat $n_y$, and fixed galaxy biases,so that Eq. \eqref{eq:limber_approx2} is exact as that the only quantity in $f$ which depends on redshift is $\xi$. Finally we use a Gaussian distribution for $n_y(z)$ centered in $2.45$ with $\sigma=0.2$, to illustrate typical approximation from using Eq. \eqref{eq:limber_approx2} for non-flat $n(z)$. 
The comparison of these three approximations with the exact results (left term of Eq. \ref{eq:limber_approx}) is reported in Fig. \ref{fig:limber_approx}.

On the auto-correlation panel, we can identify three regimes.  For large-scale-small-bin and small-scale-large-bin the approximation is valid, whereas there is a transition between these two cases with deviations at the 10-20$\%$ level. It is indeed possible to understand and predict these three regimes using simple correlation scales arguments, but we do not report them here for conciseness. 
For the cross-correlation, we can see that when the $n_y(z)$ is flat, the approximation is excellent for all cases. This is because in this case, there is no -bin border effect- and the only approximation is to assume the average over $z_1$ of $\xi_{\rm m}(z_1,z_2)$ is $\xi_{\rm m}(z_i,z_2)$ which is the case as $\xi$ is mostly linear over reasonable $\Delta z$.
Now, introducing a Gaussian distribution centered at $z=2.45$ for $n_y(z)$, we observe significant deviations with many regimes (from under-prediction at $10\%$ level to over-prediction at $20\%$ level). The transition in regimes for small $\Delta z$ has to do with the top hat bin being centred at $2.5$ and the Gaussian at $2.45$ so that the goodness of $n_y(z)=n_y(z=2.5)$ depends strongly on $\Delta z$.  For large bin, a constant $n_y(z=2.5)$ is a bad approximation for the distribution across the bin, hence large overprediction. {An explanation of this trend is provided in Section~\ref{sec: theory}: at fixed $\Delta z$ and $\theta$, the comoving width of the bin becomes smaller, while the comoving transverse scale becomes larger, as redshift increases up to $z\sim3$.}

 Thus in this work we use the left term of Eq. \eqref{eq:limber_approx}, without additional assumptions as they may fall.

\subsection{The impact of RSD effect}


\begin{figure*}
    \centering
    \includegraphics[width=0.8\linewidth]{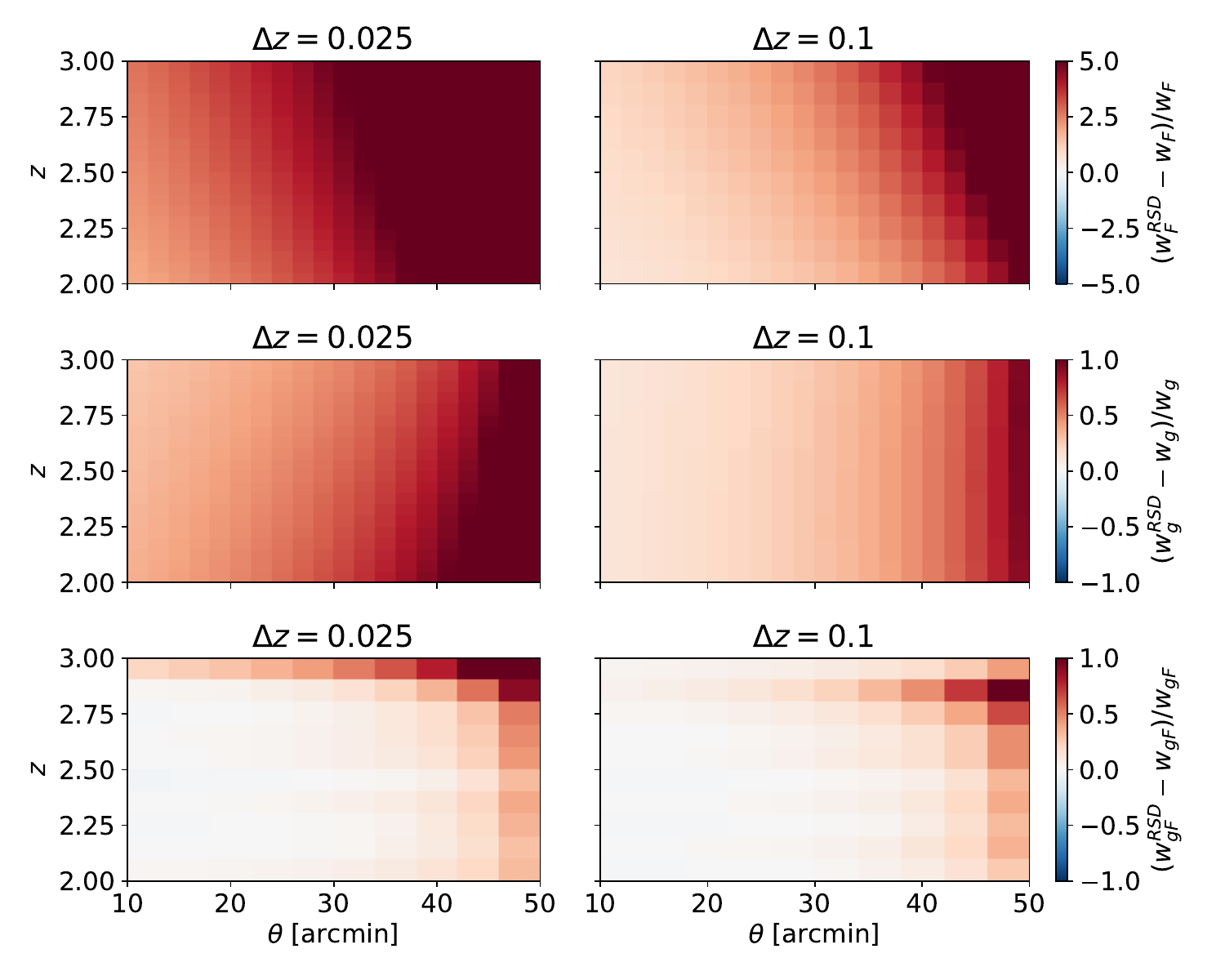}
    \caption{The fractional difference of correlation functions with and without modeling RSD effects for \lya auto-correlations (top row), galaxy auto-correlations (middle row), and \lya-galaxy cross-correlations (bottom tow). The left and right columns show two redshift bin width, $\Delta z=0.025$ and $\Delta z=0.1$, respectively.}
    \label{fig: rsd effect}
\end{figure*}

In this section, we take Eq.~\ref{eq: w2} and compare the difference by using a real-space or redshift-space correlation function. We do this for both cross- and auto-correlations, with two sets of redshift bin sizes: $\Delta z=0.025$ and $\Delta z=0.1$.

Figure.~\ref{fig: rsd effect} shows the fractional difference between the \lyaf forest auto-correlation (top row), galaxy auto-correlation (middle row), and the \lya-galaxy cross-correlation (bottom row) computed using the redshift- and real-space correlation functions, as a function of angular scales and redshifts. We see $\mathcal{O}(10)$ difference in the \lyaf auto-correlation, and $\mathcal{O}(1)$ difference in the galaxy auto-correlation, highlighting the importance of RSD effects at $z>2$. The cross-correlation is impacted at $10-50\%$ level, increasing at larger angular scales. These tests highlight that it is necessary to consider RSD effects into the models, and that the conventional $n(z)$ estimator in Eq.~\ref{eq: nu 2} is no longer applicable here.

%% file: Likelihood_william.tex
\section{Generalized and fast inference framework}
\label{sec: nz inference general}

In this section, we generalize our inference framework to general the case were: (i) we have non-linear models of $n(z)$, and (ii) we account for the photometric galaxy bias redshift evolution. We show that we can get rid of all the integrals during the inference, aiming at speeding up the process. This is similar to the linear algebra framework introduced in \cite{2025arXiv251023565D}, but it includes RSD and does not use the Limber approximation.

For the redshift distribution, let's assume we have a model for the photometric $n(z)$ which depends (linearly or not) on a set of $\boldsymbol{\theta}$ parameters (e.g. shift-stretch). The trick is to have a basis of redshift function $K_m(z)$, so that any function can be (approximately) decompose  \textit{linearly} over this basis, 
\begin{equation}
     n(z,\,\boldsymbol{\theta})\approx\sum_i f_i(\boldsymbol{\theta})\, K_i(z).
\end{equation}
This projection can be done very efficiently numerically.  Typically one can take a set of disjoint top-hat bins, with $\Delta z=0.0   1$. 
For galaxy bias, we introduce a \textit{linear} parametrization of the redshift evolution of the galaxy bias,
\begin{equation}
    b_{\rm g}(z,\mathbf{b})=\sum_j b_j\,B_j(z),
\end{equation}
where $B$ can be polynomials, power laws, the growth function etc. What is important is that this decomposition is linear. Still the redshift evolution of the bias can be highly non-linear. In practice, we should always be able to do that. 
Then
\begin{align}
 \bar w_{{\rm g}I} = & \sum_i\sum_jf_i\,b_j\underbrace{\int \diff   z^\prime ~ K_i(z^\prime)\,B_j(z^\prime) \,\bar{w}_{I}^B(z^\prime)}_{=:Q^I_{ij}} \nonumber\\
 & + \sum_if_i\underbrace{\int \diff z^\prime ~ K_i(z^\prime)\,\bar{w}_{I}^{f\mu^2}(z^\prime)}_{=:r^I_i}~, \\
 &= \mathbf{f}^\top\cdot Q^I\cdot \mathbf{b}+\mathbf{f}^\top\cdot\mathbf{r}^I
\end{align}
where $\bar{w}_{I}^{f\mu^2}$ and $\bar{w}_{I}^B$ were defined in Eq. \ref{eq: wb, wfmu}. $Q^I$ is a matrix, and $r^I$ is a vector, that are evaluated for every $I$ before running the inference, so that every step of the chain only involves linear algebra.